\documentclass[aps,preprint,superscriptaddress]{revtex4-1}
\usepackage{graphicx,amsmath,amsfonts}
\begin{document}
\title{Microscopic origin of ferromagnetism in trihalides CrCl$_3$ and CrI$_3$}
\author{Omar Besbes}
\affiliation{Faculty of Sciences, University Tunis El Manar, 2092 Tunis, Tunisia}
\author{Sergey Nikolaev}
\affiliation{International Center for Materials Nanoarchitectonics,
National Institute for Materials Science, 1-1 Namiki, Tsukuba,
Ibaraki 305-0044, Japan}
\author{Igor Solovyev}
\email{SOLOVYEV.Igor@nims.go.jp}
\affiliation{International Center for Materials Nanoarchitectonics,
National Institute for Materials Science, 1-1 Namiki, Tsukuba,
Ibaraki 305-0044, Japan}
\affiliation{Department of Theoretical Physics and Applied Mathematics, Ural Federal University,
Mira str. 19, 620002 Ekaterinburg, Russia}
\date{\today}

\date{\today}
\begin{abstract}
Microscopic origin of the ferromagnetic (FM) exchange coupling in two Cr trihalides, CrCl$_3$ and CrI$_3$, their common aspects and differences, are investigated on the basis of density functional theory combined with realistic modeling approach for the analysis of interatomic exchange interactions. For these purposes, we perform a comparative study based on the pseudopotential and linear muffin-tin orbital methods by treating the effects of electron exchange and correlation in generalized gradient approximation (GGA) and local spin density approximation (LSDA), respectively. The results of ordinary band structure calculations are used in order to construct the minimal tight-binding type models describing the behavior of the magnetic Cr $3d$ and ligand $p$ bands in the basis of localized Wannier functions, and evaluate the effective exchange coupling ($J_{\rm eff}$) between two Cr sublattices employing four different technique: (i) Brute force total energy calculations; (ii) Second-order Green's function perturbation theory for infinitesimal spin rotations of the LSDA (GGA) potential at the Cr sites; (iii) Enforcement of the magnetic force theorem in order to treat both Cr and ligand spins on a localized footing; (iv) Constrained total-energy calculations with an external field, treated in the framework of self-consistent linear response theory. We argue that the ligand states play crucial role in the ferromagnetism of Cr trihalides, though their contribution to $J_{\rm eff}$ strongly depends on additional assumptions, which are traced back to fundamentals of adiabatic spin dynamics. Particularly, by neglecting ligand spins in the Green's function method, $J_{\rm eff}$ can easily become antiferromagnetic, while by treating them as localized, one can severely  overestimate the FM coupling. The best considered approach is based on the constraint method, where the ligand states are allowed to relax in response to each instantaneous reorientation of the Cr spins, controlled by the external field. Furthermore, the differences of the electronic structure of Cr trihalides in GGA and LSDA, and their impact on the exchange coupling are discussed in details, as well as the possible roles played by the on-site Coulomb repulsion $U$.
\end{abstract}

\maketitle
\section{\label{sec:Intro} Introduction}
\par Cr trihalides, Cr$X_3$ (where $X=$ Cl, Br, or I), and other van der Waals compounds with the rhombohedral structure have attracted recently much attention due to discovery of two-dimensional (2D) ferromagnetism~\cite{CrGeTe3_Nature,CrI3_Nature}, which has tremendous importance for the development of ultra-compact spintronics. Furthermore, there is a general fundamental interest in microscopic origin of this phenomenon. Basically, there are two main question: (i) the existence of ferromagnetic (or any other long-range magnetic) order at finite temperatures in the 2D case, which formally contradicts to Mermin-Wagner theorem~\cite{MerminWagner}; (ii) the origin of ferromagnetic (FM) interactions themselves. The first restriction is known to be removed by anisotropic interactions~\cite{BrunoPRL2001}, as was recently confirmed for van der Waals magnets~\cite{CrGeTe3_Nature,CrI3_Lui,CrI3_Lado}. Regarding the second issue, the ferromagnetism of Cr trihalides is basically a bulk property. For instance, bulk samples of CrBr$_3$ and CrI$_3$ are known to be ferromagnetic with the Curie temperature of $33$ K and $68$ K, respectively~\cite{Tsubikawa1960,Bene1969}, while CrCl$_3$ is a layered antiferromagnet (i.e., also consisting of FM layers) with the N\'eel temperature of $17$ K~\cite{Bene1969}.

\par Then, why are Cr trihalides ferromagnetic? An easy answer would be the following: Cr is trivalent, located in the octahedral environment, and the Cr-$X$-Cr angle is close to $90^\circ$. Therefore, the Cr-Cr coupling is expected to be ferromagnetic according to Goodenough-Kanamori-Anderson (GKA) rules~\cite{Kanamori_GKA}, mainly due to intraatomic exchange interaction responsible for Hund's first rule at the ligand ($X$) sites. However, the GKA rule for the $90^\circ$-degree exchange is not very conclusive, as it relies on specific paths and processes for the exchange coupling, while in reality there can be several competing mechanisms supporting either ferromagnetism or antiferromagnetism, so that the final answer strongly depends on the ratio of relevant physical parameters~\cite{Chaloupka}. In fact, Kanamori himself admitted that there are exceptions from this rule in the $90^\circ$-degree case~\cite{Kanamori_GKA}, and in our work we will show that the exchange coupling in Cr trihalides can easily become antiferromagnetic, depending on the considered model.

\par Another important issue is how these GKA rules work in practical first-principles calculations based on density functional theory (DFT), and how to map properly this or any other theory of interacting electrons onto effective spin model and not to lose the contributions that can be responsible for the GKA rules~\cite{JHeisenberg,Stepanov,review2008,CrO2PRB2015}. The problem is indeed very complex and there are many tricky issues in it. Particularly, the mapping onto the spin model frequently assumes the local character of exchange-correlation (xc) interactions, following a similar concept of local spin density approximation (LSDA), which often supplements DFT calculations and is also ``local''. However, the main FM mechanism of the $90^\circ$ exchange is basically non-local: the magnetic $3d$ states spread from two Cr sites to the neighboring ligand site and there interact via Hund's rule coupling~\cite{Kanamori_GKA}, which has the same origin as the canonical direct exchange interaction proposed by Heisenberg almost a centaury ago~\cite{Heisenberg}. In this respect, it is interesting to note that CrBr$_3$  was indeed regarded as a rare example of materials (together with CrO$_2$) where ferromagnetism arises from the direct exchange interaction~\cite{Yosida}.

\par There is quite a common tendency in the electronic structure calculations: if (typically due to additional approximations) such calculations underestimate the FM coupling, the missing effect is ascribed to the direct exchange~\cite{CrO2PRB2015,Ku,Mazurenko2007}. Nevertheless, this step looks somewhat irrational, because the effect of direct exchange interactions is already included in LSDA, which is used as a starting point for many electronic structure calculations, even despite it is formally ``local''. Furthermore, LSDA also includes the important effects of screening for the direct exchange interactions due to electron correlations. Indeed, the xc energy in LSDA is given by
\noindent
\begin{equation}
E_{\rm xc}^{\rm LSDA}[n, \boldsymbol{m} ] = \int d\boldsymbol{r}  n(\boldsymbol{r}) \varepsilon_{\rm xc} [n(\boldsymbol{r}), | \boldsymbol{m}(\boldsymbol{r}) | ]
\label{eq:xcLSDA}
\end{equation}
($n$ and $\boldsymbol{m}$ being total electron and spin magnetization density, respectively). Thus, at each point, the xc energy is given by $n$ and $\boldsymbol{m}$ \emph{at the same point}, which is the definition of locality of the LSDA functional. However, $n$ and $\boldsymbol{m}$ can be equivalently expanded in terms of a complete set of localized Wannier functions for the occupied states~\cite{WannierRevModPhys}. In this representation, $\boldsymbol{m}(\boldsymbol{r})$ is the sum of Wannier function contributions coming from \emph{different} magnetic sites. Therefore, although $E_{\rm xc}^{\rm LSDA}$ is local with respect to the total magnetization density, it includes non-local effects caused by overlap of individual contributions to $\boldsymbol{m}$ centered at different magnetic sites in the Wannier representation.

\par In this work we will address this problem in details by considering two characteristic examples of Cr trihalides: CrCl$_3$ and CrI$_3$. Particularly, how should one calculate the interatomic exchange interactions within DFT in order to take into account all relevant processes, responsible for the FM coupling in the $90^\circ$ case? To this end, we will argue that it may not be enough to consider the reorientation of only transition-metal spins (for instance, within the commonly used Green's function perturbative approach for infinitesimal spin rotations~\cite{JHeisenberg}), as there is a large contribution to the exchange energy associated with the ligand sites~\cite{Oguchi}, which should be also taken into account. Nevertheless, it does not automatically guarantee the ``right'' solution of the problem because the answer strongly depends on how the ligand spins are treated, tracing back to fundamentals of adiabatic spin dynamics. For instance, by enforcing the magnetic force theorem in order to treat the Cr and ligand spins on the same localized footing, one can severely overestimate the FM coupling. It turns out that the best considered approach is to fully relax the ligand states for each instantaneous configuration of the Cr spins: it correctly describes the FM coupling, both in CrCl$_3$ and CrI$_3$, and yields the effective exchange coupling constant comparable with results of brute force total energy calculations. We will show how this problem can be efficiently solved in the framework of self-consistent linear response theory~\cite{SCLR}.

\par The rest of the paper is organized as follows. In Sec.~\ref{sec:Method} we will describe our method based on construction and analysis of the minimal tight-binding type models for the Cr $3d$ and ligand $p$ bands of CrCl$_3$ and CrI$_3$ in the basis of localized Wannier functions. All our calculations are based on either LSDA or generalized gradient approximation (GGA) as implemented in, respectively, the linear muffin-tin orbital (LMTO) and pseudopotential Quantum-ESPRESSO (QE) methods. The on-site Coulomb repulsion $U$ should not play a decisive role in the origin of ferromagnetism in Cr trihalides, as argued in Appendix~\ref{sec:AppendixA}. Then, in Sec.~\ref{sec:Jij} we will consider different approaches for calculations of effective intersublattice exchange coupling: the Green's function technique (Sec.~\ref{sec:GF}), magnetic force theorem (Sec.~\ref{sec:MF}), and constrained calculations with an external magnetic field (Sec.~\ref{sec:constraint}). Finally, in Sec.~\ref{sec:conc}, we will give a brief summary on our work.

\section{\label{sec:Method} Method}
\par
\begin{figure}[tbp]
\begin{center}
\includegraphics[width=10cm]{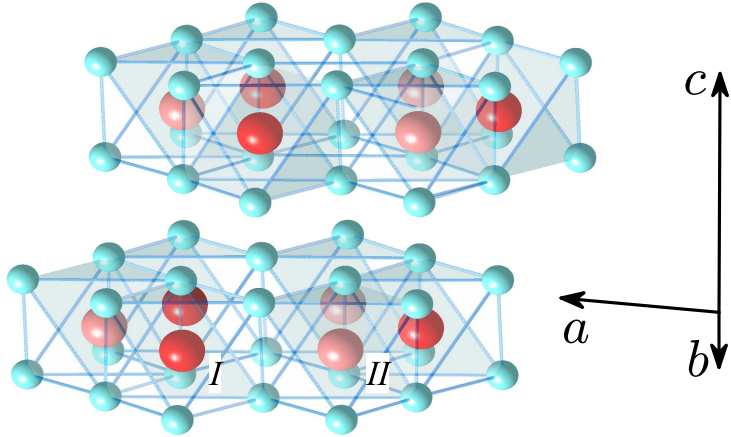}
\end{center}
\caption{(Color online)
Fragment of the crystal structure of CrCl$_3$. Cr and Cl atoms are denoted by large and small spheres respectively. Cr atoms belonging to different sublattices are shown by different colors and numbered as `$I$' and `$II$'.}
\label{fig.str}
\end{figure}

 We use the $R \overline{3}$ structure for both CrCl$_3$ and CrI$_3$ (shown in Fig.~\ref{fig.str}) with the experimental parameters reported in Refs.~\cite{CrCl3str} and \cite{CrI3str}, respectively. As for the electronic structure calculations, we perform a comparative study based on the LMTO method in the atomic-spheres approximation (ASA)~\cite{LMTO} and the pseudopotential QE method realized in the plane-wave basis~\cite{qe}. Furthermore, all LMTO calculations have been performed in LSDA, while in QE we employed the generalized gradient approximation (GGA) with the Perdew-Burke-Ernzerhof exchange-correlation (xc) functional~\cite{PBE}. We use the mesh of $8 \times 8 \times 8$ ($10 \times 10 \times 10$) ${\bf k}$-points in the Brillouin zone in the case of QE (LMTO) and the energy cutoff of $110$ Ry in the QE calculations.

\par
\begin{figure}[tbp]
\begin{center}
\includegraphics[width=7cm]{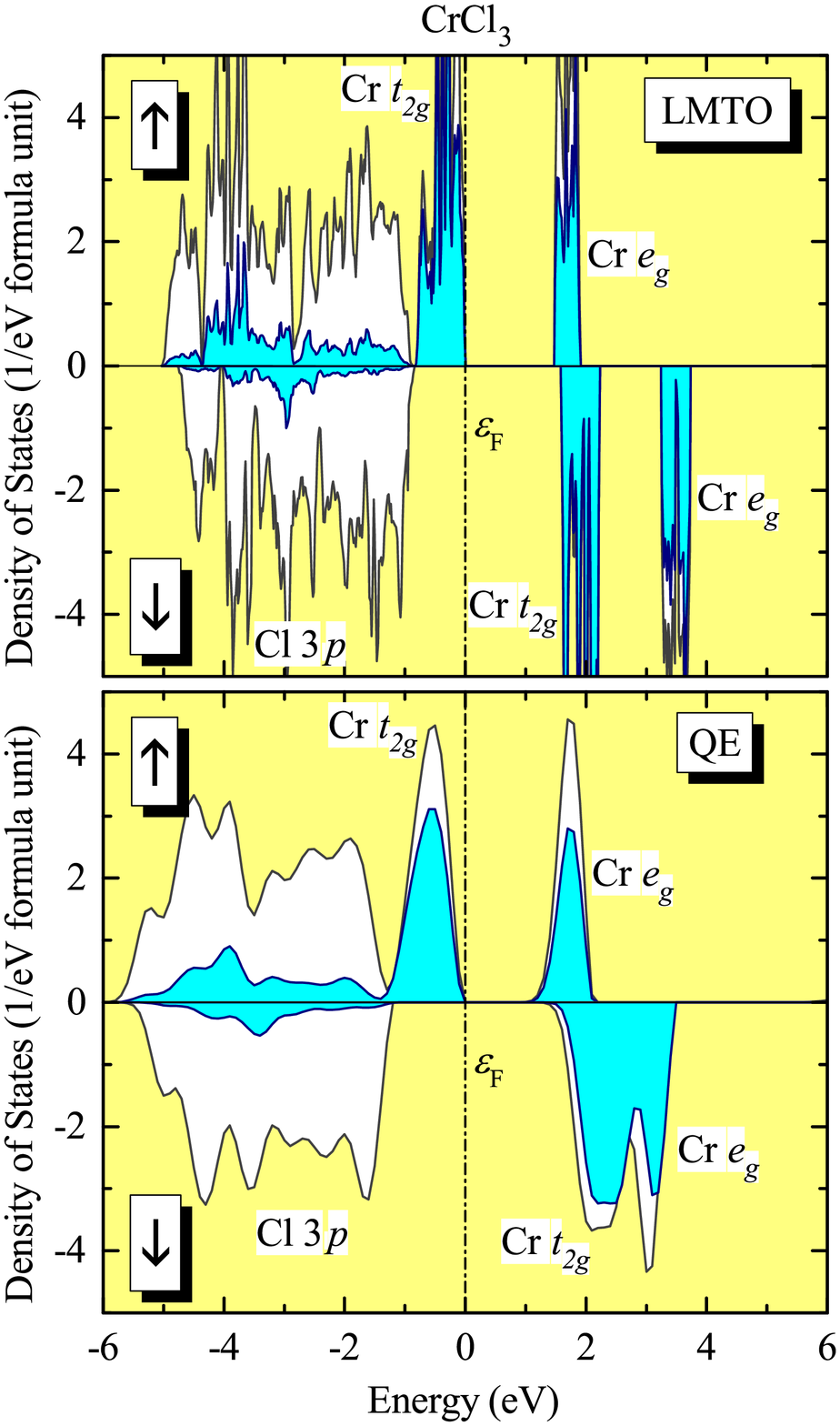}
\includegraphics[width=7cm]{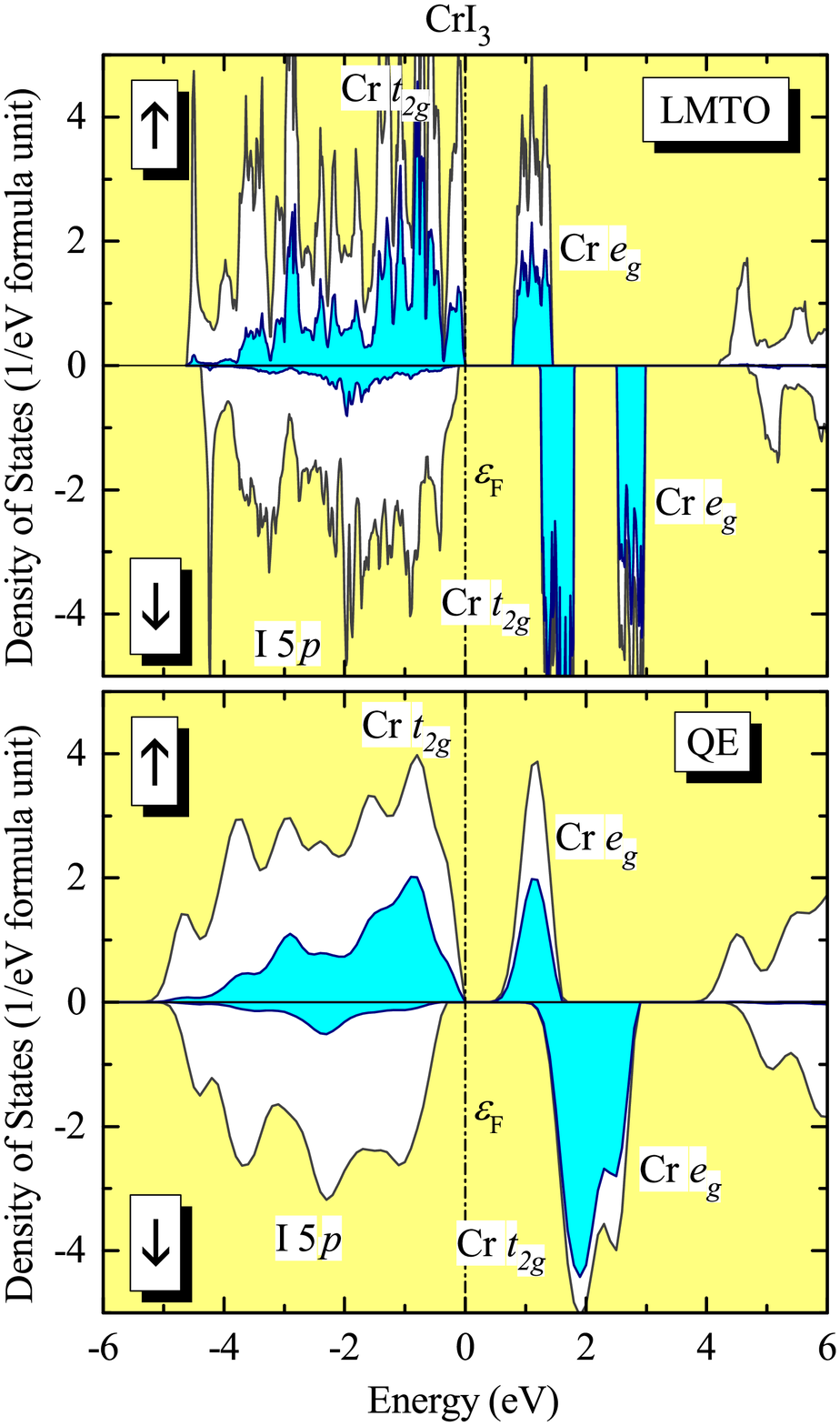}
\end{center}
\caption{(Color online)
Total and partial densities of states of CrCl$_3$ (left) and CrI$_3$ (right) as obtained in LMTO (top) and QE (bottom) methods, supplemented with LSDA and GGA, respectively.
The shaded area shows contributions of the Cr $3d$ states.
Positions of the main bands are indicated by symbols. The Fermi level is at zero energy (shown by dot-dashed line).}
\label{fig.DOS}
\end{figure}

 Corresponding electronic structure, obtained for the FM state, is shown in Fig.~\ref{fig.DOS}.
\noindent Owing to the octahedral environment of the Cr sites, the magnetic $3d$ states are split into the low-lying $t_{2g}$ ($t$) and high-lying $e_{g}$ ($e$) groups. In CrCl$_3$, the Cr $3d$ bands are well separated from the Cl $3p$ one, so that even for the majority ($\uparrow$) spin channel there is a finite gap between these two groups of bands. Therefore, in this case one can consider two types of models: the $dp$ one, which explicitly includes the $3d$ states of Cr as well as the $p$ states of the ligand atoms, and the pure $d$ one, consisting only of the Cr $3d$ states. Nevertheless, in CrI$_3$, due to the additional upward shift of the I $5p$ states, the $\uparrow$-spin Cr $t_{2g}$ band merges into the I $5p$ one. Therefore, in this case we are able to consider only the more general $dp$ model. The electronic structure of CrCl$_3$ and CrI$_3$ obtained in the QE method is in excellent agreement with the results of previous GGA calculations~\cite{Wang2011}. The differences between LMTO and QE results will be discussed below.

\par Next, we construct the Wannier functions spanning the subspace of the Cr $3d$ and ligand $p$ bands and define the model $dp$ Hamiltonian, $\hat{H}^{\uparrow, \downarrow} = [H_{ij}^{ab}]^{\uparrow, \downarrow}$, separately for the majority ($\uparrow$) and minority ($\downarrow$) spin states, as the matrix elements of the original LSDA (GGA) Hamiltonian in the basis of Wannier functions, which are numbered by indices $a$ and $b$ at sites $i$ and $j$. In order to construct the Wannier functions, we use the projector operator method and the maximally localized Wannier functions (MLWF) technique (implemented in the Wannier90 package) in the case of, respectively, LMTO and QE~\cite{review2008,WannierRevModPhys,wannier90}. For the trial orbitals in LMTO we use basis functions for the Cr $3d$ and ligand $p$ states. The completeness of the Wannier basis guarantees that the obtained band structure in the region of the Cr $3d$ and ligand $p$ bands fully coincides with the original LSDA (GGA) one. After that, the $d$ model for CrCl$_3$ was constructed by starting from the $dp$ model and eliminating the Cr $3p$ band by means of the projector operator technique, both in LMTO and QE methods.

\par
\begin{figure}[tbp]
\begin{center}
\includegraphics[width=10cm]{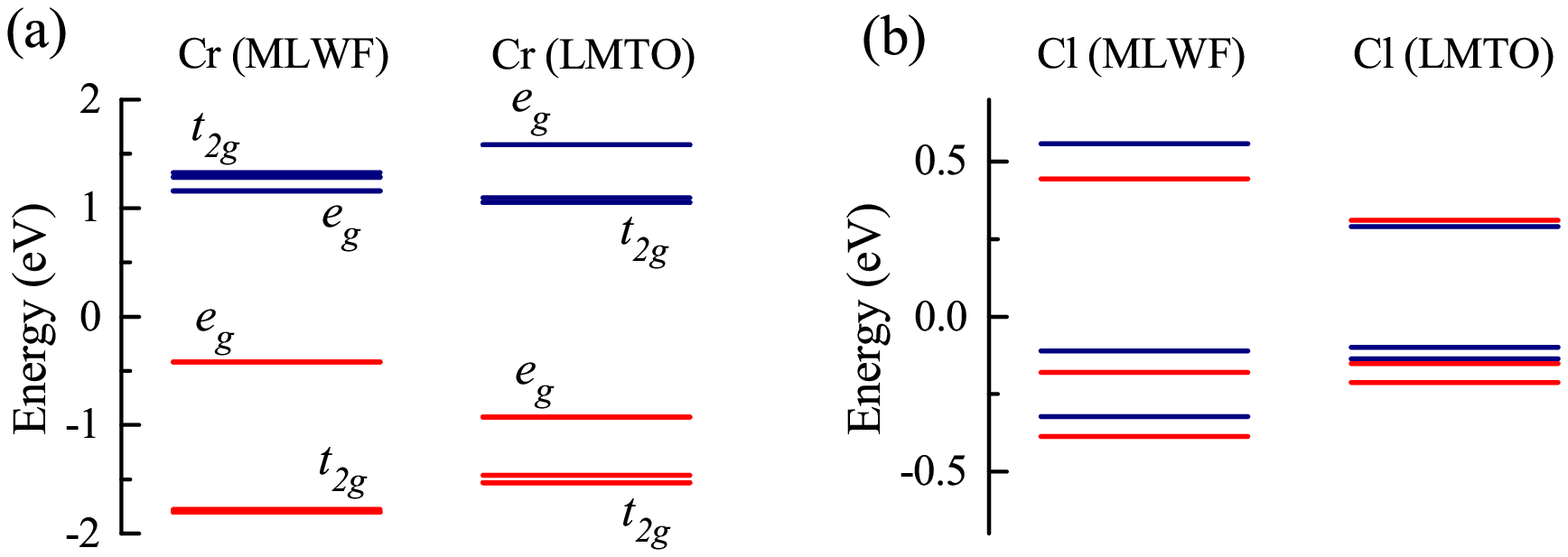} \\
\includegraphics[width=10cm]{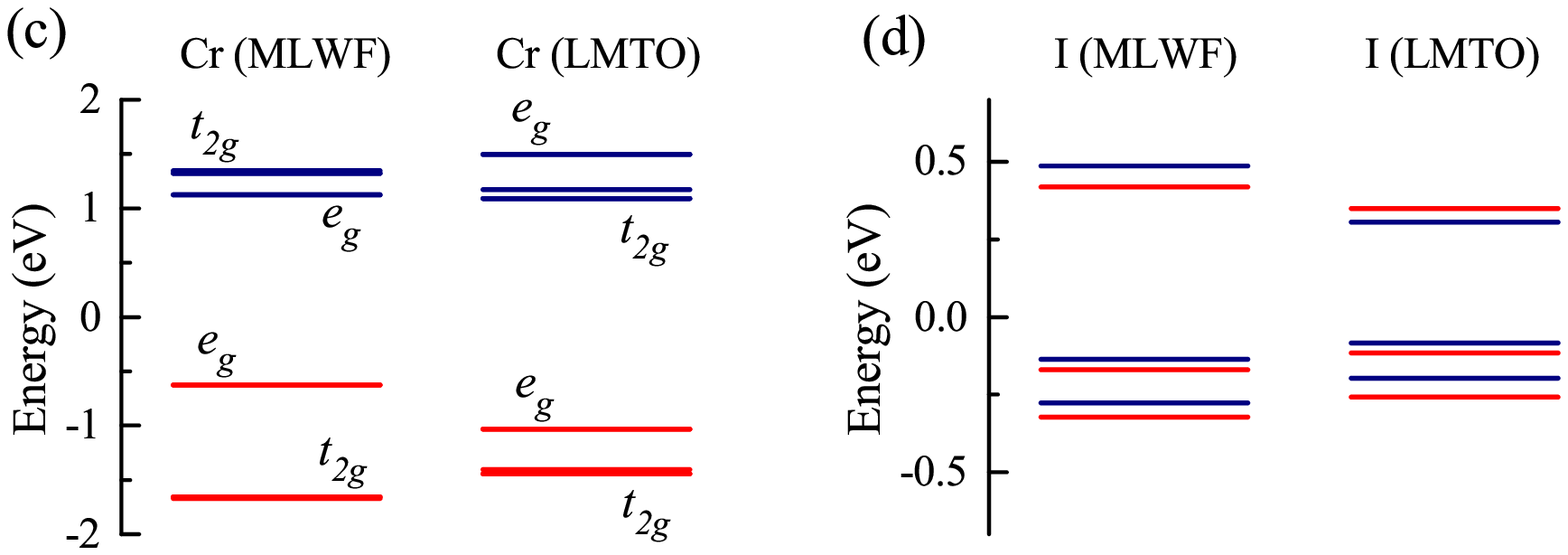}
\end{center}
\caption{(Color online)
Atomic level splitting in the $dp$ model for CrCl$_3$ (top) and CrI$_3$ (bottom) as obtained in LMTO method and maximally localized Wannier functions (MLWF) technique, which is based on the QE method: (a,c) Cr $3d$ states and (b,d) ligand $p$ states. The majority
and minority spin states are denoted by red (light) and blue (dark) colors, respectively.}
\label{fig.CF}
\end{figure}

 The atomic level splitting, which is obtained by diagonalizing the site-diagonal part of $\hat{H}^{\uparrow, \downarrow}$, is shown in Fig.~\ref{fig.CF}.
The $\uparrow$-$\downarrow$ spin splitting of the Cr $3d$ states, driven by the xc interactions in LSDA (GGA), is larger than the $t_{2g}$-$e_{g}$ crystal-field splitting within each projection of spin. In the ASA-LMTO method, the $\uparrow$-$\downarrow$ level splitting is nearly rigid and well described by the phenomenological expression $I_{\rm Cr} M_{\rm Cr}$, where $I_{\rm Cr} \approx 0.85$ eV is the Stoner parameter and $M_{\rm Cr} = 3 \mu_{\rm B}$ is the local spin moment of Cr$^{3+}$, as expected in LSDA~\cite{Gunnarsson}. In the QE (MLWF) method, however, this simple picture is no longer valid and there is an additional splitting between the $t_{2g}$ and $e_{g}$ orbitals, which also depends on spin. Particularly, the $t_{2g}$-$e_{g}$ splitting for spin $\downarrow$ is substantially smaller than that for spin $\uparrow$, which reflects similar behavior of the density of states in Fig.~\ref{fig.DOS}. Nevertheless, this is to be expected because the QE method takes into account the asphericity of the Kohn-Sham potential~\cite{lda}, which is totally ignored in ASA. Furthermore, this asphericity is additionally amplified in GGA (in comparison with LSDA). The spin splitting of the ligand $p$ levels is small in comparison with the crystal field splitting within each projection of spin. The latter effect is again larger in the QE calculations.

\par
\begin{figure}[tbp]
\begin{center}
\includegraphics[width=10cm]{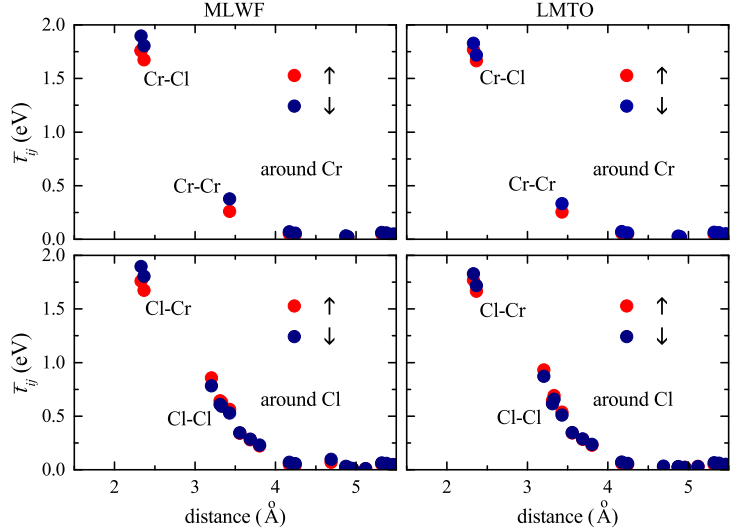}
\end{center}
\caption{(Color online)
Distance-dependence of averaged transfer integrals around the Cr and Cl sites in the $dp$ model for CrCl$_3$ as obtained in the LMTO method and maximally localized Wannier functions (MLWF) technique based on the QE method.}
\label{fig.tCrCl3}
\end{figure}

 Fig.~\ref{fig.tCrCl3} shows the behavior of averaged transfer integrals, $\bar{t}_{ij} = \sqrt{\sum_{ab} H_{ij}^{ab} H_{ji}^{ba}}$, for CrCl$_3$.
\noindent Despite substantial difference in details of QE and LMTO calculations, the obtained parameters are very similar in two considered methods. The transfer integrals around Cr sites are basically limited to the neighboring Cr-Cl and Cr-Cr bonds. The transfer integrals around ligand sites involve more ligand-ligand bonds. Nevertheless, all transfer integrals are short-ranged and practically vanish at the distance $d \sim 4.5$ \AA. Very similar tendency was found for CrI$_3$, which is not shown here.

\par
\begin{figure}[tbp]
\begin{center}
\includegraphics[width=4.5cm]{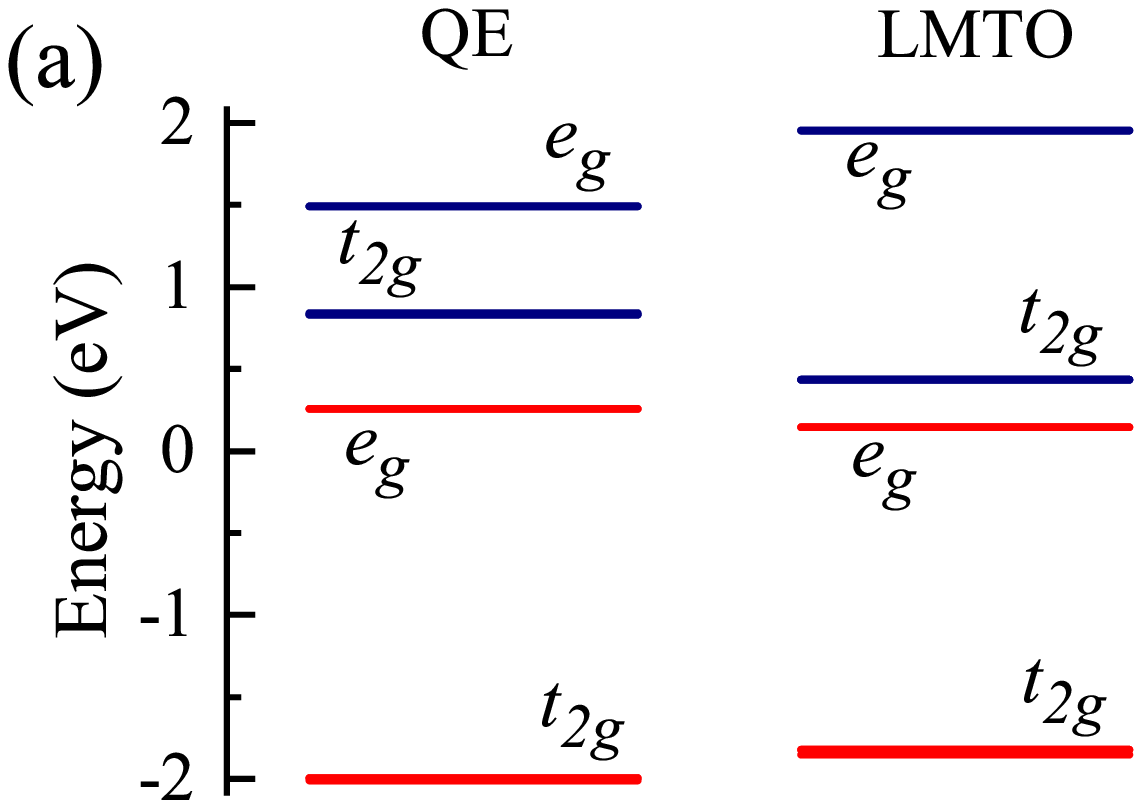}
\end{center}
\begin{center}
\includegraphics[width=10cm]{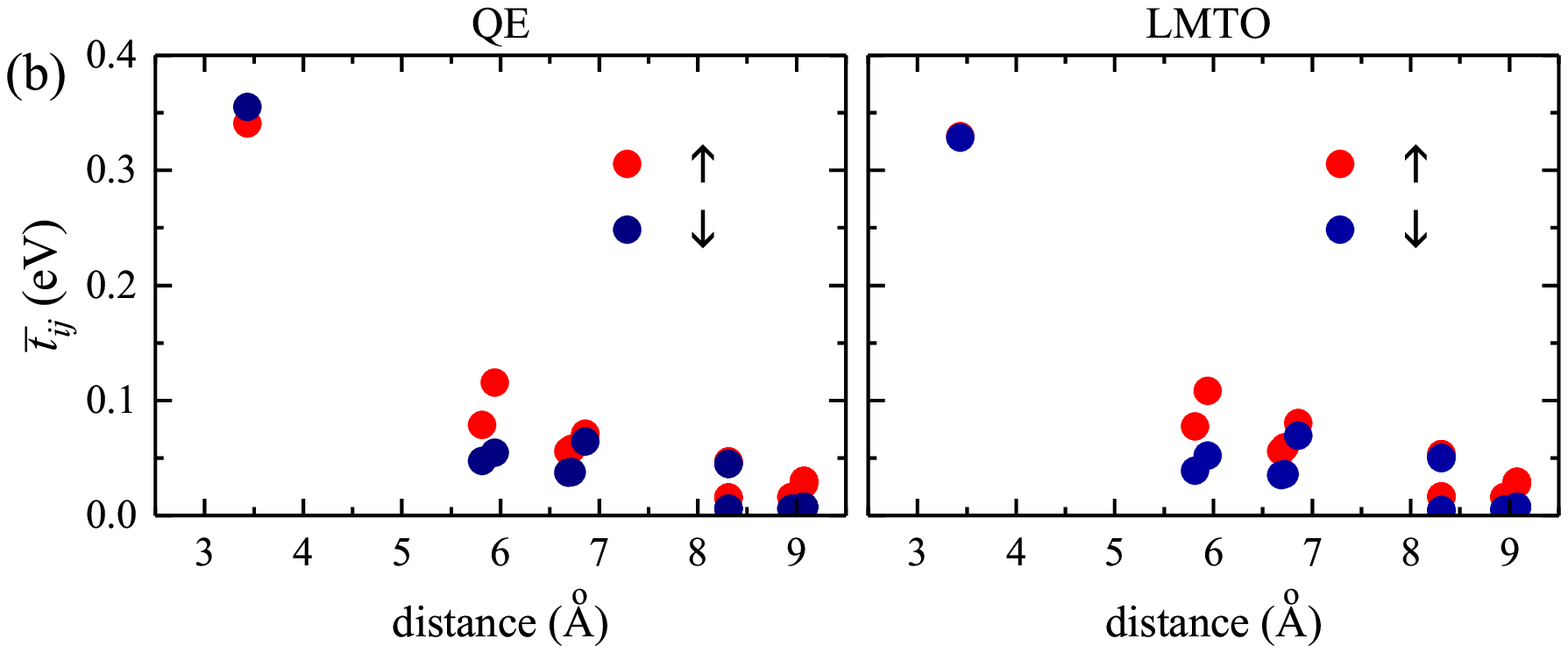}
\end{center}
\caption{(Color online)
Results of $d$ model for CrCl$_3$ as obtained in QE and LMTO methods (see text for details of the construction of the model): (a) atomic level splitting of the Cr $3d$ states and (b) distance-dependence of averaged transfer integrals around Cr sites.}
\label{fig.CrCl3d}
\end{figure}

 Parameters of the $d$ model for CrCl$_3$ are summarized in Fig.~\ref{fig.CrCl3d}.
\noindent The elimination of the Cl $3p$ bands considerably increases the crystal-field splitting between the $t_{2g}$ and $e_{g}$ orbitals, being consistent with an old conjecture by Kamanori that the $10Dq$ splitting results mainly from the hybridization of the transition-metal $3d$ states with the ligand $p$ states~\cite{Kanamori}. Nevertheless, the structure of the atomic Cr $3d$ levels remains quite different in LMTO and QE (MLWF). As we shall see below, this difference strongly affects the behavior of nearest-neighbor (nn) exchange interactions. As expected, the transfer integrals become more long-ranged in comparison with the ones in the $dp$ model~\cite{PRB06}. However, in all other respects their behavior is very similar in LMTO and QE.

\section{\label{sec:Jij} Magnetic interactions}
\par
\begin{figure}[tbp]
\begin{center}
\includegraphics[width=8cm]{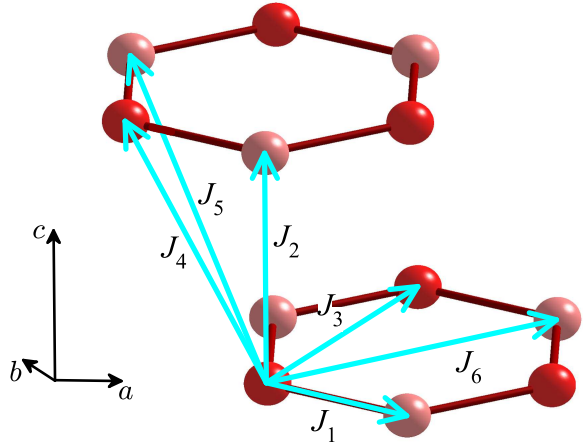}
\end{center}
\caption{(Color online)
Notations of exchange interactions in CrCl$_3$ and CrI$_3$ up to six nearest neighbors. Atoms belonging to different Cr sublattices are denoted by different colors.}
\label{fig.Ji}
\end{figure}

 In this Section we explore abilities of different models and techniques for the evaluation of interatomic exchange interactions in CrCl$_3$ and CrI$_3$. Our ultimate goal is to elucidate the microscopic origin of the FM coupling in these compounds and we will show that the problem is indeed very nontrivial. The basic idea is to map the total energy difference associated with the reorientation of the Cr spins onto the model
\noindent
\begin{equation}
{\cal H}_S = -\frac{1}{2} \sum_{ij} J_{j} \boldsymbol{e}_{i} \cdot \boldsymbol{e}_{i+j},
\label{eqn:Heisenberg}
\end{equation}
\noindent where $\boldsymbol{e}_{i}$ is the \emph{direction} of spin at site $i$, and $J_{j}$ is the corresponding exchange coupling between (central) site $i$ and the one located at the lattice point $i+j$. Unless specified  otherwise, we will concentrate on the behavior of the effective coupling between two Cr sublattices $I$ and $II$ (see Fig.~\ref{fig.Ji}): $J_{\rm eff} = \sum_{j \in II} J_{j}$ (provided that the central site belongs to the  sublattice $I$). If interactions with other Cr sites, except the nn ones, are negligibly small, we will have a trivial relation $J_{\rm eff} \approx J_{1}$, which holds approximately for CrCl$_3$ and CrI$_3$.
\noindent

\par The most straightforward estimate for $J_{\rm eff}$ is to use the total energy difference between the antiferromagnetic (AFM) and FM states in the QE method, $\Delta E = E_{\rm AFM} - E_{\rm FM}$, which yields $21.35$ and $34.93$ meV per one formula unit of CrCl$_3$ and CrI$_3$, respectively, corresponding to $J_{\rm eff} = \Delta E/3 = 7.12$ and $11.64$ meV. This result clearly indicates that the coupling is indeed ferromagnetic and stronger in the case of CrI$_3$. Thus, this result can be used as a reference for other models and approximations.

\subsection{\label{sec:GF} Green's function method}
\par The first model we consider is based on the Green's function method~\cite{JHeisenberg}. The main idea here is to define the xc field $\hat{\boldsymbol{b}}_{i} = (0,0, \hat{b}_{i}^{z})$, associated with an arbitrary site $i$, where $\hat{b}_{i}^{z} = \hat{H}_{ii}^\uparrow - \hat{H}_{ii}^\downarrow$, and consider the energy change caused by infinitesimal rotations of this field, $\hat{\boldsymbol{b}}_{i} \rightarrow \hat{\boldsymbol{b}}_{i} = (\theta_{i} , 0, 1$$-$$\theta_{i}^{2}/2 )\hat{b}_{i}^{z}$ ($\theta_{i}$ being the azimuthal angle), as a perturbation. Then, to 2nd order in $\theta$, the corresponding energy change will be given by Eq.~(\ref{eqn:Heisenberg}) with the parameters
\noindent
\begin{equation}
J_i = \frac{1}{2\pi} {\rm Im} \int_{- \infty}^{\varepsilon_{\rm F}} d \varepsilon \, {\rm Tr}_L \left\{
\hat{b}_{0}^{z} \hat{G}_{0i}^{\uparrow}(\varepsilon)
\hat{b}_{i}^{z} \hat{G}_{i0}^{\downarrow}(\varepsilon)
\right\},
\label{eqn:Jij}
\end{equation}
\noindent where $\hat{G}^{\uparrow, \downarrow}_{0i}(\varepsilon) = [ \varepsilon - \hat{H}^{\uparrow, \downarrow} ]^{-1}_{0i}$ is the one-electron Green function between sites $0$ and $i$, $\varepsilon_{\rm F}$ is the Fermi energy, and ${\rm Tr}_L$ denotes the trace over orbital indices~\cite{JHeisenberg}. Importantly, Eq.~(\ref{eqn:Jij}) describes the change of single-particle (Kohn-Sham) energies for the occupied states, while other contributions to the total energy are supposed to cancel out in the 2nd order of $\theta$ as a result of the magnetic force theorem~\cite{JHeisenberg}. This technique is widely used in the LMTO method where, owing to the pseudoatomic basis, one can easily define local magnetic moments at each site. The advantage of the Wannier basis is that it allows us to transfer this technique to other methods, including those originally formulated in the basis of plane waves, as in the QE method.

\par In principle, Eq.~(\ref{eqn:Jij}) can also be used for the calculation of the exchange coupling between Cr and ligand sites. Nevertheless, we begin here with the standard procedure and employ this equation only for the Cr-Cr interactions, while assuming that the main role of the intermediate ligand states is to assist these interactions by connecting two Cr sites. The results of these calculations for $J_{1}$ are summarized in Table~\ref{tab:J1}.
\noindent
\begin{table}[h!]
\caption{Nearest-neighbor exchange interaction ($J_{1}$, in meV) in the $dp$ model for CrCl$_3$ and CrI$_3$ as obtained in the Green's function method for infinitesimal rotations of the Cr $3d$ spins (results of the Cr $3d$ model for CrCl$_3$ are shown for comparison in parentheses). The model was constructed in the LMTO method and using maximally localized Wannier functions technique (MLWF) in the QE method.}
\label{tab:J1}
\begin{ruledtabular}
\begin{tabular}{ccc}
 & CrCl$_3$ & CrI$_3$ \\
\hline
MLWF  &           $-1.63$           ($-2.75$) & $2.37$ \\
LMTO  & $\phantom{-}4.07$ ($\phantom{-}2.47$) & $2.98$ \\
\end{tabular}
\end{ruledtabular}
\end{table}
\noindent The LMTO and QE methods provide quite consistent description for $J_{1}$ in the case of CrI$_3$, where the FM character of this interaction is consistent with experimental data. Nevertheless, the situation is totally different in CrCl$_3$, where even the sign of $J_{1}$ is different in LMTO and QE, thus putting the latter data in clear disagreement with the experiment. This is the first puzzle we need to solve and below we will argue that such behavior is related to the electronic structure of CrCl$_3$ and CrI$_3$ in LMTO and QE.

\par Since the $dp$ and $d$ models provide qualitatively similar results and encounter the same problem for CrCl$_3$ ($J_{1}>0$ in LMTO, while $J_{1}<0$ in QE), it is convenient to start our analysis with the simplest $d$ model. An analytical expression for $J_{1}$ can be found by employing superexchange theory~\cite{PWA,KugelKhomskii}, which relies on additional approximations but provides some insight into behavior of $J_{1}$. It yields
\noindent
\begin{equation}
J_{1} \approx \left( \frac{1}{\Delta^{\uparrow \uparrow}_{te}} - \frac{1}{\Delta^{\uparrow \downarrow}_{te}} \right) t^{2}_{te} - \frac{1}{\Delta^{\uparrow \downarrow}_{tt}} t^{2}_{tt},
\label{eqn:SE}
\end{equation}
\noindent where $t^{2}_{te} = \sum_{ a \in t, b \in e } \left( H_{01}^{ab} \right)^{2}$, $t^{2}_{tt} = \sum_{a, b \in t} \left( H_{01}^{ab} \right)^{2}$, and $\Delta$ is the intraatomic energy splitting between the $t_{2g}$ and either $e_{g}$ ($te$) or $t_{2g}$ ($tt$) levels with the same ($\uparrow \uparrow$) or opposite ($\uparrow \downarrow$) directions of spin. Then, it is clear that smaller $\Delta^{\uparrow \uparrow}_{te}$ and larger $\Delta^{\uparrow \downarrow}_{te}$ in LMTO (see Fig.~\ref{fig.CrCl3d}a) favor the FM coupling. On the other hand, smaller $\Delta^{\uparrow \downarrow}_{tt}$, also in LMTO, will tend to stabilize the AFM coupling. However, since $t^{2}_{tt} < t^{2}_{te}$ ($t^{2}_{te} = 0.044$ eV$^2$, while $t^{2}_{tt} = 0.028$ eV$^2$, according to MLWF calculations), the last term plays a less important role. These tendencies clearly explain why $J_{1} > 0$ in LMTO, while $J_{1} < 0$ in QE. One of the lessons learned from this analysis is that the FM coupling in the $d$ model can be stabilized only if $\Delta^{\uparrow \uparrow}_{te}$ is relatively small. This is the main reason why the ferromagnetism totally disappears, even in LMTO, if one tries to correct this superexchange picture by adding the on-site Coulomb repulsion, which additionally increases $\Delta^{\uparrow \uparrow}_{te}$, as discussed in Appendix~\ref{sec:AppendixA}. Similar conclusion can be arrived by considering the so-called ``double exchange'' (DE) contribution to $J_{1}$, obtained in the 2nd order expansion for $\hat{G}_{i0}^{\downarrow}(\varepsilon)$ with respect to $( \hat{b}_{0}^{z} )^{-1} = ( \hat{b}_{i}^{z} )^{-1}$ in Eq.~(\ref{eqn:Jij})~\cite{PRL99,CrO2PRB2015}, which yields
\noindent
\begin{equation}
J_{1}^{\rm DE} = \frac{1}{2\pi} {\rm Im} \int_{- \infty}^{\varepsilon_{\rm F}} d \varepsilon \, {\rm Tr}_L \left\{
\hat{G}_{01}^{\uparrow}(\varepsilon) \hat{H}_{10}^{\downarrow}
\right\}.
\label{eqn:DE}
\end{equation}
\noindent In comparison with Eq.~(\ref{eqn:SE}), this expression takes into account the contributions proportional to $1/\Delta^{\uparrow \uparrow}$ and neglects the contributions proportional to $1/\Delta^{\uparrow \downarrow}$, assuming that the latter are small. As expected~\cite{PRL99}, $J_{1}^{\rm DE}$ is ferromagnetic and can be estimated in LMTO and QE as $18.85$ and $16.23$ meV, respectively. Thus, we clearly observe again the excess of ferromagnetism in the case of LMTO, which is closely related to the structure of intraatomic level splitting.

\par Then, why does not this problem occur in CrI$_3$? Regarding the intraatomic level splitting, the situation is pretty much similar in CrCl$_3$ and CrI$_3$ (see Fig.~\ref{fig.CF}): in both cases, the LMTO and QE methods provide two different schemes of atomic level splitting (but with clear similarity between CrCl$_3$ and CrI$_3$, if one compares separately the results of either LMTO or QE calculations). Nevertheless, the LMTO and QE methods  provide rather consistent description for $J_{1}$ in CrI$_3$ (see Table~\ref{tab:J1}). The reason is again related to the electronic structure of CrI$_3$, which can be classified as the charge-transfer insulator~\cite{ZSA}: the $\uparrow$-spin I $5d$ and Cr $3d$ bands strongly overlap so that these two groups of states become entangled with each other (see Fig.~\ref{fig.DOS}). In such situation, the distribution of the Cr $3d$ states over the energy, which mainly controls the value of $J_{1}$ in Eq.~(\ref{eqn:Jij}), is determined by the $dp$ hybridization, while the atomic level splitting is less important. This naturally explains similarity of LMTO and QE data for CrI$_3$.

\par
\begin{figure}[tbp]
\begin{center}
\includegraphics[width=7cm]{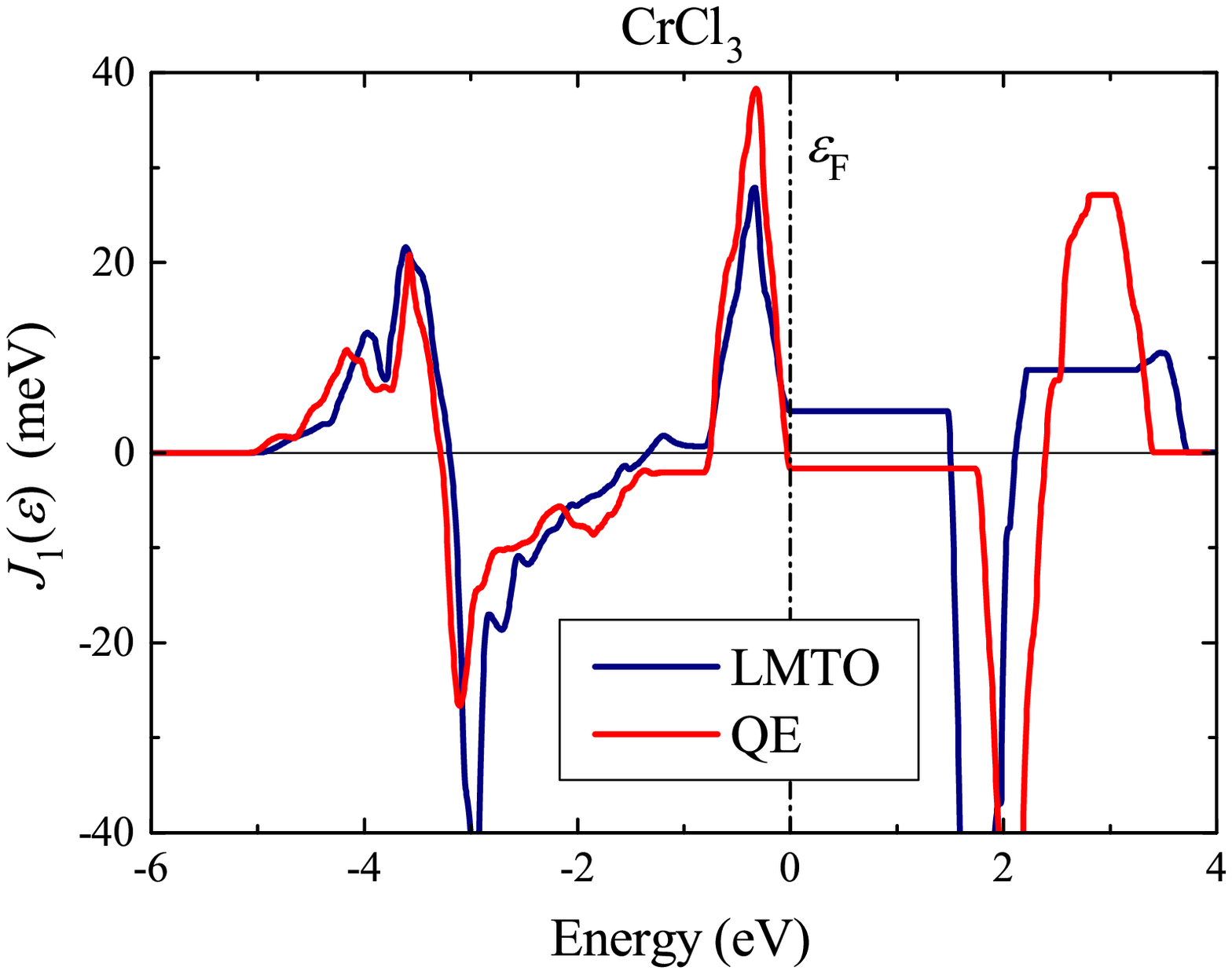}
\includegraphics[width=7cm]{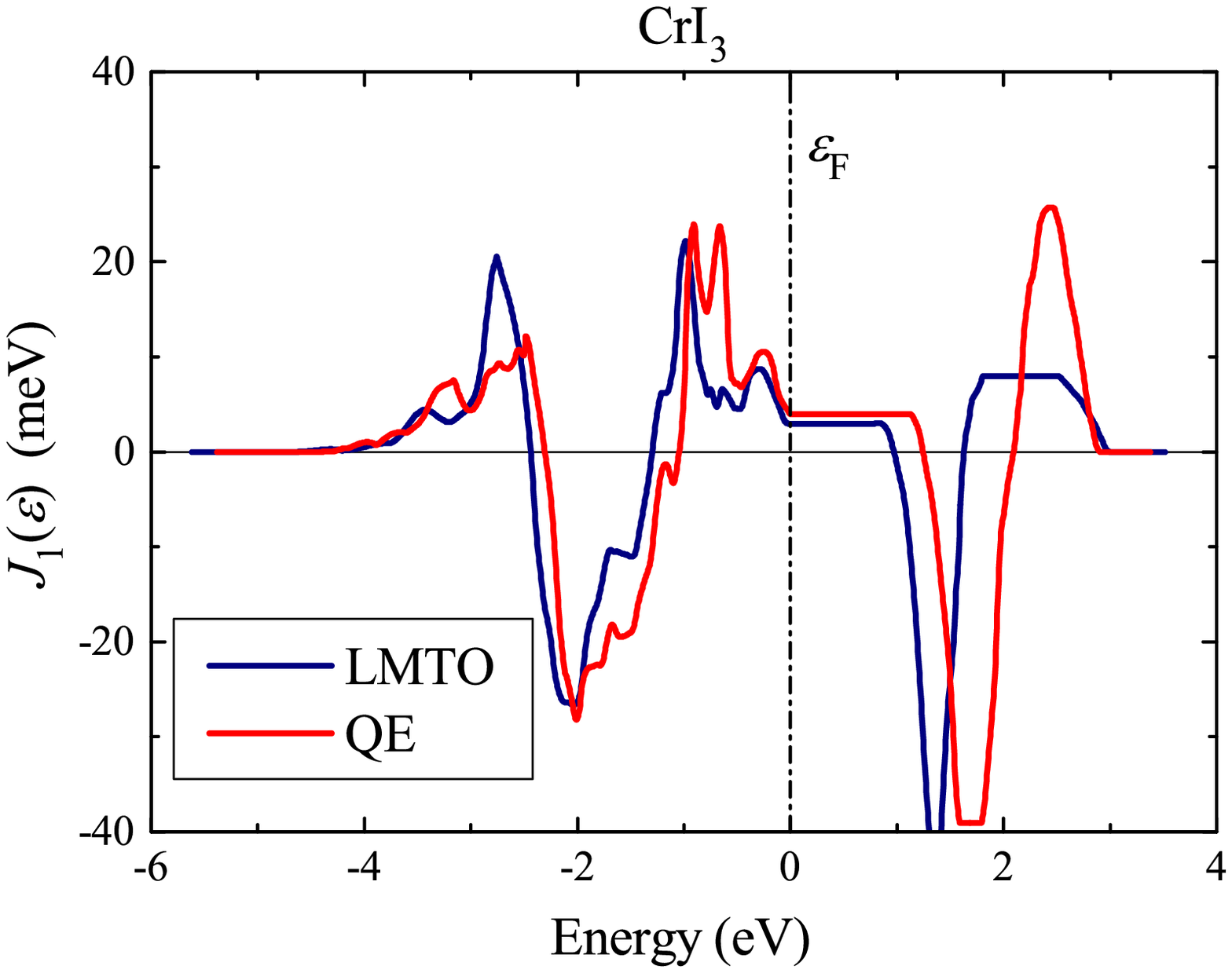}
\end{center}
\caption{(Color online)
Band-filling dependence of nearest-neighbor exchange interaction in the case of CrCl$_3$ (left) and CrI$_3$ (right). The true Fermi level, corresponding to the nominal number of valence electrons, is at zero energy (shown by dot-dashed line).}
\label{fig.J1e}
\end{figure}

 In order to further elucidate the impact of the electronic structure on interatomic exchange interactions in CrCl$_3$ and CrI$_3$, it is instructive to consider the band-filling dependence of $J_{1}$, which is again calculated using Eq.~(\ref{eqn:Jij}), but varying the position of the Fermi level (\ref{fig.J1e}).
\noindent
\noindent In CrCl$_3$, $J_{1}(\varepsilon)$ practically vanishes after integration over the Cl $3p$ band, spreading from $-5$ to $-1$ eV (see Fig.~\ref{fig.DOS}), even though there is a substantial variation of $J_{1}(\varepsilon)$ within the band. Thus, the contribution of the Cl $3p$ band to $J_{1}$ is small, that justifies the use of the $d$ model for these purposes, as was done above. Then, in the QE method, the shape of $J_{1}(\varepsilon)$ in the Cr $t_{2g}$ band (spreading from $-1$ to $0$ eV) is nearly symmetric relative to the band center so that the total contribution of this band to $J_{1}$ vanishes. This means that the FM and AFM interactions originating from virtual excitations from the Cr $t_{2g}$ band to unoccupied states cancel each other. In the LMTO method, the shape of $J_{1}(\varepsilon)$ in the Cr $t_{2g}$ region is slightly deformed, leading to some asymmetry, which is in turn responsible for the appearance of uncompensated FM interactions, as was discussed above. In CrI$_3$, however, the I $5p$ and Cr $t_{2g}$ states form a common band. The magnetic interactions in this band obey some general principles: namely, the ferromagnetism is expected at the beginning and the end of the band filling, while the antiferromagnetism is expected in the middle of the band filling~\cite{Heine1,Heine2}. Obviously, if the band were fully isolated, we would have $J_{1}(\varepsilon_{\rm F})=0$. Nevertheless, the interaction with unoccupied states makes in ferromagnetic, both in LMTO and QE calculations.

\par Parameters of the main exchange interactions, spreading up to six nearest neighbors (see Fig.~\ref{fig.Ji}), are reported in Table~\ref{tab:Ji}. One can see that the nn interaction clearly dominates, so that $J_{\rm eff} \approx J_{1}$ holds for CrCl$_3$ and to lesser extent for CrI$_3$. Furthermore, using the obtained parameters $J_{i}$ and employing Tyablikov's random-phase approximation~\cite{tyab}, the Curie temperature for CrI$_3$ can be estimates as $83$ K, which is in fair agreement with the experimental value of $68$ K~\cite{Bene1969}.
\noindent
\begin{table}[h!]
\caption{Parameters of the exchange interaction (in meV) as obtained using the QE (MLWF) method, supplemented with the Green's function method for the infinitesimal rotations of the Cr $3d$ spins. Notations of exchange interactions are shown in Fig.~\ref{fig.Ji}.}
\label{tab:Ji}
\begin{ruledtabular}
\begin{tabular}{lcccccc}
          & $J_{1}$           & $J_{2}$           & $J_{3}$ & $J_{4}$ & $J_{5}$ & $J_{6}$ \\
\hline
CrCl$_3$  & $-1.63$           & $-0.10$           & $0.21$  & $0.09$  & $0.07$  & $-0.28$ \\
CrI$_3$   & $\phantom{-}2.37$ & $\phantom{-}0.25$ & $0.84$  & $0.46$  & $0.68$  & $-0.26$ \\
\end{tabular}
\end{ruledtabular}
\end{table}

\par Summarizing this part, the Green's function based expression, Eq.~(\ref{eqn:Jij}), correctly describes the FM coupling in CrI$_3$, but not in CrCl$_3$. The FM character of exchange interactions obtained for CrCl$_3$ in the ASA-LMTO method is probably fortuitous as it does not properly treat the asphericity of the Kohn-Sham potential, as was demonstrated by more rigorous full-potential QE calculations. Thus, we are still missing some important mechanism that stabilizes the FM ground state in CrCl$_3$ (and probably plays some role also in CrI$_3$). In the remaining part of this section we will go beyond Eq.~(\ref{eqn:Jij}) for the analysis of interatomic exchange interactions and argue that missing FM contributions are related to the ligand $p$ states, though the answer strongly depends on how these contributions are treated.

\subsection{\label{sec:MF} Enforcement of the magnetic force theorem}
\par First, we note that the expression~(\ref{eqn:Jij}) suffers from a systematic error: although it follows from the magnetic force theorem, which allows us to replace the total energy change by the change of single-particle energies, it formally violates this theorem. The magnetic force theorem can be proven if a certain rotation of the xc field $\hat{\boldsymbol{b}}_{i} \rightarrow \tensor{R}_{\theta_{i}} \hat{\boldsymbol{b}}_{i}$ rotates the spin magnetization by the same angle: $\hat{\boldsymbol{m}}_{i} \rightarrow \tensor{R}_{\theta_{i}} \hat{\boldsymbol{m}}_{i}$. Note that in these notations, $\hat{\boldsymbol{b}}_{i}$ is the vector $\hat{\boldsymbol{b}}_{i} = (\hat{b}_{i}^{x},\hat{b}_{i}^{y},\hat{b}_{i}^{z})$, where each of its components is the matrix in the Wannier basis, while  $\tensor{R}_{\theta_{i}}$ is the $3 \times 3$ tensor acting on $\hat{b}_{i}^{x}$, $\hat{b}_{i}^{y}$ , and $\hat{b}_{i}^{z}$. Then, since the xc energy is invariant under spin rotations, the total energy change is described by the single-particle part given by Eq.~(\ref{eqn:Jij})~\cite{PRB98}. This property holds in LSDA, unrestricted Hartree-Fock approximation, and for any xc functional satisfying the condition of gauge invariance~\cite{PRB98}. However, the rotation $\hat{\boldsymbol{b}}_{i} \rightarrow \tensor{R}_{\theta_{i}} \hat{\boldsymbol{b}}_{i}$ (which is typically used in practical calculations) does not necessarily lead to the rotation $\hat{\boldsymbol{m}}_{i} \rightarrow \tensor{R}_{\theta_{i}} \hat{\boldsymbol{m}}_{i}$. More specifically, considering an iterative process inherent to solving the Kohn-Sham equations, the rotation $\hat{\boldsymbol{b}}_{i} \rightarrow \tensor{R}_{\theta_{i}} \hat{\boldsymbol{b}}_{i}$ will indeed lead to the rotation $\hat{\boldsymbol{m}}_{i} \rightarrow \tensor{R}_{\theta_{i}} \hat{\boldsymbol{m}}_{i}$ at the \emph{input} of iteration. However, at the \emph{output} of the same iteration, the magnetization will tend to relax to the ground state configuration and additionally rotate away from $\tensor{R}_{\theta_{i}} \hat{\boldsymbol{m}}_{i}$. This problem was noticed, for instance, by Stocks~\textit{et al.}~\cite{Stocks} and considered in details by Bruno~\cite{BrunoPRL2003}, who also proposed to adjust the magnetization by applying a constraining field fixing the direction of the spin magnetization along $\tensor{R}_{\theta_{i}} \hat{\boldsymbol{m}}_{i}$ for a given rotation of the xc field, $\hat{\boldsymbol{b}}_{i} \rightarrow \tensor{R}_{\theta_{i}} \hat{\boldsymbol{b}}_{i}$. Below, we closely follow this strategy, also using for these purposes the linear response theory~\cite{SCLR}.

\par Namely, let $\vec{\boldsymbol{b}}$ be the column vector composed of $\hat{\boldsymbol{b}}_{i}$ at different sites of the unit cell (including ligands), $(\vec{\boldsymbol{b}})^{T} = (\ \dots \ , \hat{\boldsymbol{b}}_{i}, \ \dots \ )$, and $\vec{\boldsymbol{m}}$ be a similar vector for the magnetization. Then, knowing the $x$ component $(\vec{m}^{x})^{T} = (\ \dots \ , \theta_{i} \hat{m}_{i}^{z}, \ \dots \ )$ of the rotated magnetization, one can find the constraining field $\vec{\boldsymbol{h}} = (\vec{h}^{x},0,0)$, which reproduces $\vec{m}^{x}$, using the linear response theory as $\vec{h}^{x} = \left[ \left( \mathcal{R}^{\uparrow \downarrow} \right)^{-1} + \left( \mathcal{R}^{\downarrow \uparrow} \right)^{-1} \right] \vec{m}^{x} - \vec{b}^{x}$, where $\mathcal{R}^{\uparrow \downarrow}$ and $\mathcal{R}^{\downarrow \uparrow}$ are the matrix elements of the response tensor, whose explicit expression can be found in Ref.~\cite{SCLR}. Note, that throughout this work we use the notations where the spin magnetization is related to the density matrix $\hat{n}$ as $\hat{\boldsymbol{m}} = {\rm Tr}_S \{ \hat{\boldsymbol{\sigma}} \hat{n} \}$ ($\hat{\boldsymbol{\sigma}}$ being the vector of Pauli matrices and ${\rm Tr}_S$ being the trace over spin variables) and the magnetic field (both constraining and xc one) enters the Kohn-Sham equations as $\frac{1}{2} \hat{\boldsymbol{h}} \cdot \hat{\boldsymbol{\sigma}}$, which holds at each site of the system.

\par Using $\vec{\boldsymbol{h}}$ and $\vec{\boldsymbol{b}}$, one can calculate the total energy corresponding to the ``rotated'' configuration of spins as is typically done in the constraint method:
\noindent
\begin{equation}
E ( \vec{\boldsymbol{m}} ) = E_{\rm sp} ( \vec{\boldsymbol{b}}+\vec{\boldsymbol{h}} ) - \frac{1}{2} {\rm Tr}_L \{ \vec{\boldsymbol{h}} \cdot \vec{\boldsymbol{m}} \},
\label{eqn:constraint}
\end{equation}
\noindent where $E_{\rm sp} ( \vec{\boldsymbol{b}}+\vec{\boldsymbol{h}} )$ is the sum of the occupied Kohn-Sham energies calculated for the field $\vec{\boldsymbol{b}}+\vec{\boldsymbol{h}}$, and $\vec{\boldsymbol{h}} \cdot \vec{\boldsymbol{m}}$ stands for the dot product of two vectors with the summation over atomic sites. For $\vec{\boldsymbol{h}} = 0$, Eq.~(\ref{eqn:constraint}) is reduced to Eq.~(\ref{eqn:Jij}).

\par
\begin{figure}[tbp]
\begin{center}
\includegraphics[width=10cm]{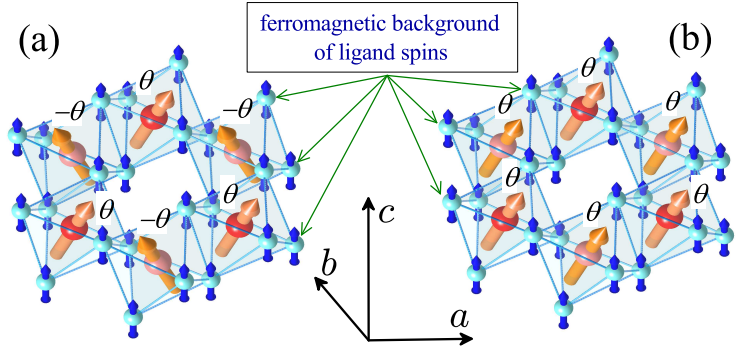}
\end{center}
\caption{(Color online)
Magnetic geometries used in calculations of the effective exchange coupling between Cr sublattices. (a) The Cr spins were rotated ``antiferromagnetically'' by $\theta$ and $-$$\theta$ ($\theta$ being the azimuthal angle). (b) The spins were rotated ``ferromagnetically'' by the same angle $\theta$ in both sublattices. The ligand spins were fixed to have the same directions and values as in the ferromagnetic state.}
\label{fig.mgeometry}
\end{figure}

\begin{figure}[tbp]
\begin{center}
\includegraphics[width=7cm]{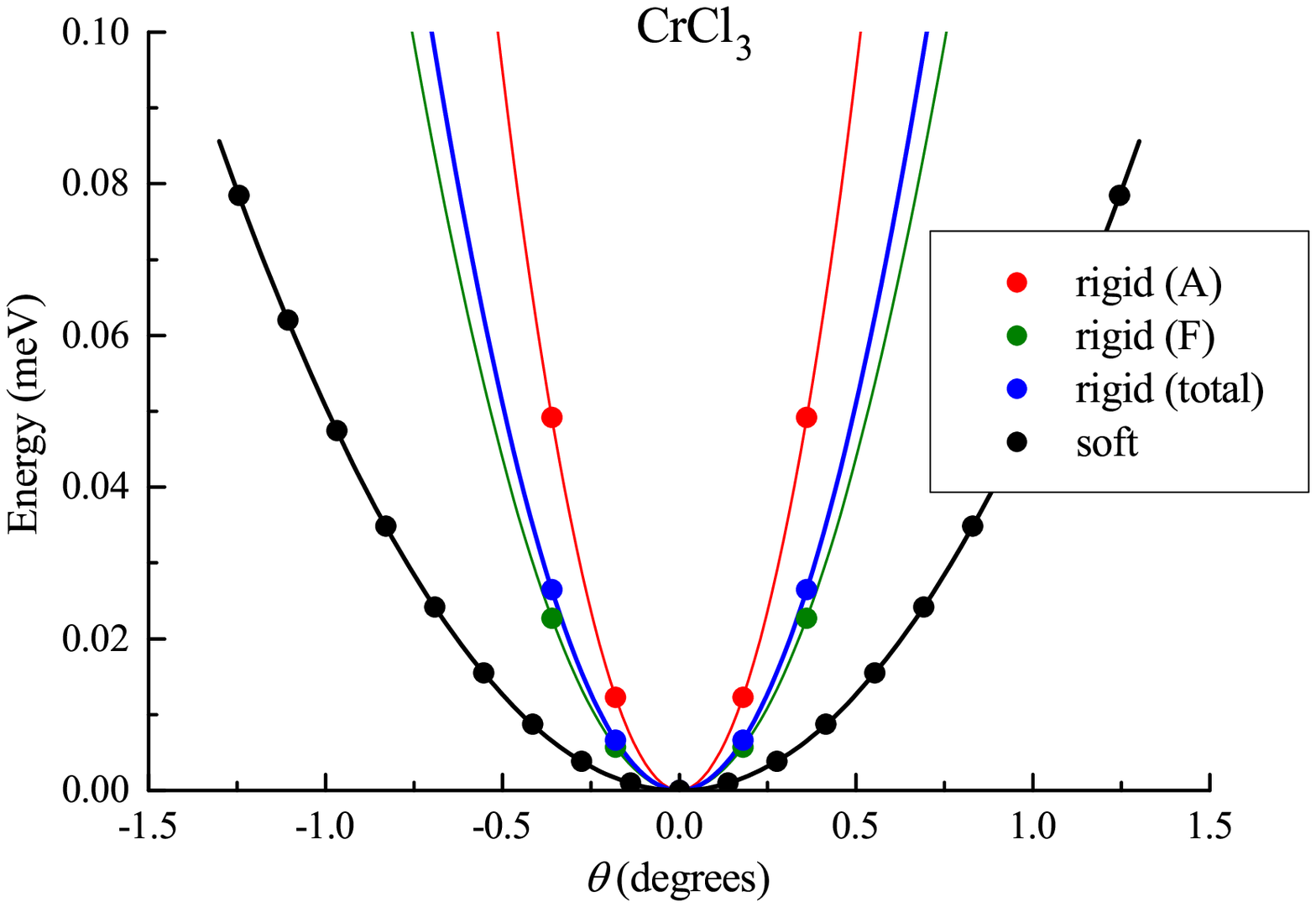}
\includegraphics[width=7cm]{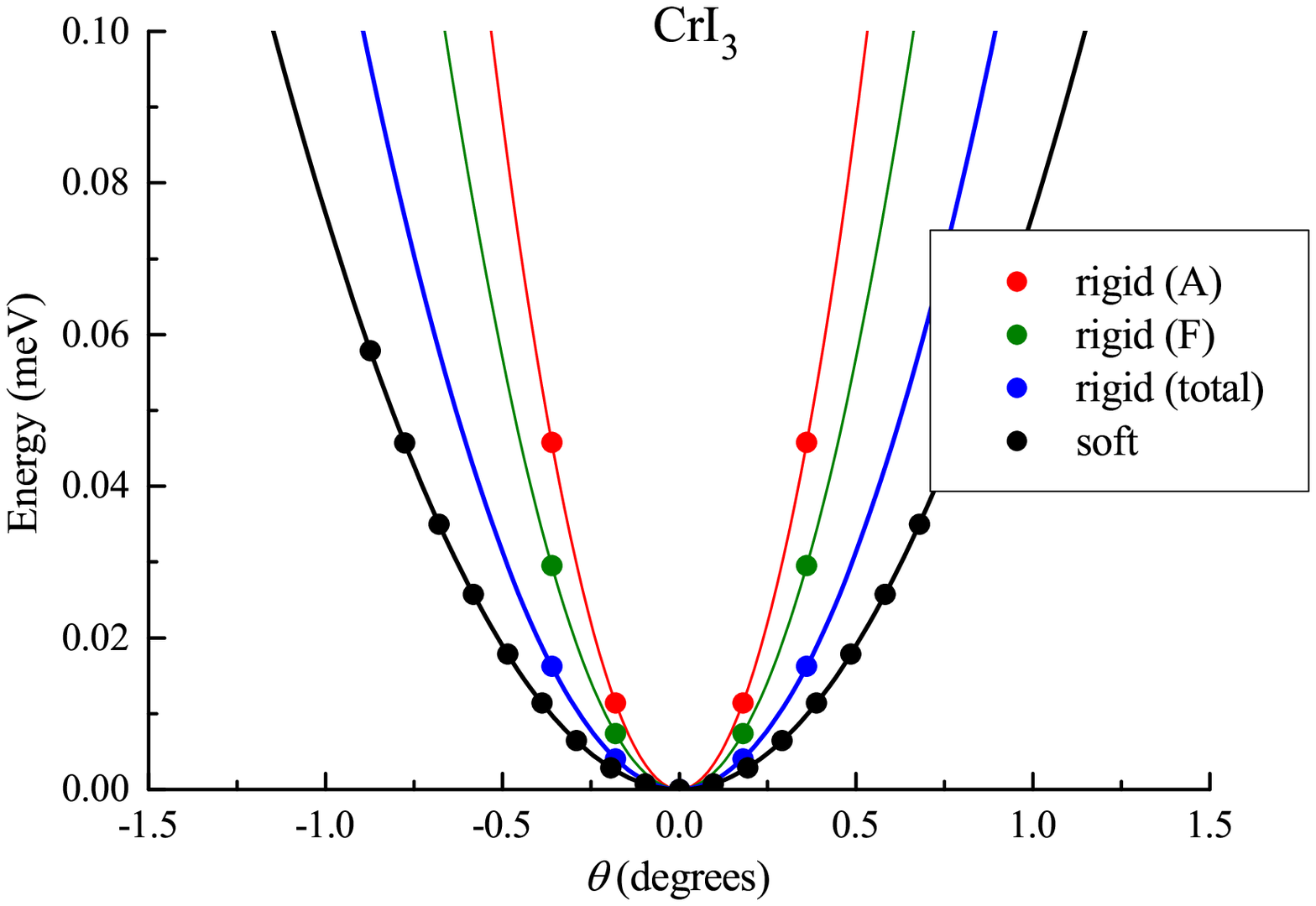}
\end{center}
\caption{(Color online)
Results of constraint calculations for the energies (per two formula units) of canted spin configurations, as obtained in the QE (MLWF) method for CrCl$_{3}$ (left) and CrI$_{3}$ (right). The ``rigid'' constraint is based on Eq.~\ref{eqn:constraint}, where the  magnetization density is constrained at all sites, including the ligands, in order to fulfil the magnetic force theorem. ``A'' and ``F'' correspond to the antiferromagnetic and ferromagnetic rotations of the Cr spins, as shown in Figs.~\ref{fig.mgeometry}a and b, respectively, and ``total'' is the difference of A and F. The ``soft'' constraint is based on Eq.~\ref{eqn:Elr}, where the directions of spins are controlled by the external field applied only at the Cr sites. The calculated values are shown by symbols, while the interpolation $E = 6 J_{\rm eff} \theta^{2}$ is shown by solid curves.}
\label{fig.e-theta}
\end{figure}

 In order to obtain $J_{\rm eff}$, we first consider the magnetic configuration shown in Fig.~\ref{fig.mgeometry}a, where the spins in two Cr sublattices are rotated by $\theta$ and $-$$\theta$ relative to the FM axis $z$. The corresponding energy  (per one unit cell including \emph{two} Cr sites) is denoted as $E(\theta,-\theta)$.
\noindent
\noindent We call these rotations ``antiferromagnetic'', referring to the alignment of transversal components of the rotated spins. Corresponding exchange coupling can be evaluated as
\noindent
\begin{equation}
J_{\rm eff}^{\rm A} = \frac{1}{3} \frac{\partial^2 E(\theta,-\theta)}{\partial (2 \theta)^2},
\label{eqn:jeff}
\end{equation}
\noindent where the prefactor $1/3$ takes into account three nn bonds in the $xy$ plane. The obtained energies are shown in Fig.~\ref{fig.e-theta} and the corresponding parameters are summarized in Table~\ref{tab:J1c}.
\noindent
\begin{table}[h!]
\caption{Parameters of the effective exchange coupling (in meV) as obtained using the QE (MLWF) method and Eq.~(\ref{eqn:constraint}) for the energies of infinitesimal rotations of spins, which enforces the magnetic force theorem. $J_{\rm eff}^{\rm A}$ and $J_{\rm eff}^{\rm F}$ are the values obtained for the antiferromagnetic and ferromagnetic rotations of spins in two Cr sublattices, as shown in Fig.~\ref{fig.mgeometry}a and b, respectively, and $J_{\rm eff} = J_{\rm eff}^{\rm A}-J_{\rm eff}^{\rm F}$ is the total coupling. The results of LMTO calculations are shown in parentheses for comparison.}
\label{tab:J1c}
\begin{ruledtabular}
\begin{tabular}{lccc}
          & $J_{\rm eff}^{\rm A}$ & $J_{\rm eff}^{\rm F}$ & $J_{\rm eff}$  \\
\hline
CrCl$_3$  &      $207.74$ ($286.10$) &        $\phantom{1}95.98$ ($145.11$)  & $111.76$                     ($140.99$) \\
CrI$_3$   &      $193.23$ ($203.14$) &        $124.63$ ($144.78$)            & $\phantom{1}68.60$ ($\phantom{1}58.36$) \\
\end{tabular}
\end{ruledtabular}
\end{table}
\noindent The constrained energy reveals perfect quadratic dependence on the angle $\theta$. Moreover, after enforcing the magnetic force theorem, Eq.~(\ref{eqn:constraint}) correctly reproduces the FM coupling for both CrI$_3$ and CrCl$_3$. However, the value of this coupling is too large compared to typical energies of magnetic excitations in transition-metal oxides and related compounds.

\par In the following, we will show that the reason is related to a very fundamental problem of how to treat the contribution of ligand sites to the exchange coupling between Cr sites. Indeed, the $dp$ hybridization induces feasible magnetization at the ligand sites (Table~\ref{tab:lm}).
\noindent
\begin{table}[h!]
\caption{Local magnetic moments (in $\mu_{\rm B}$) at the Cr and ligand sites as obtained for the $dp$ model in the QE (MLWF) method. The results of LMTO calculations are shown in parenthesis for comparison.}
\label{tab:lm}
\begin{ruledtabular}
\begin{tabular}{lcc}
          & Cr                & ligand                        \\
\hline
CrCl$_3$  & $2.951$ ($3.145$) & $\phantom{-}0.016$ ($-0.048$) \\
CrI$_3$   & $3.105$ ($3.370$) & $-0.035$           ($-0.123$) \\
\end{tabular}
\end{ruledtabular}
\end{table}
\noindent The difference between the QE (MLWF) and LMTO data are quite expected: the values of local magnetic moments depend on the crystal-field splitting, which is very different in QE and LMTO, as was discussed above (see Fig.~\ref{fig.CF}). Nevertheless, the QE and LMTO data show the same tendency: the Cr moment increases from CrCl$_3$ to CrI$_3$ (due to the increase of $dp$ hybridization and overlap of the I $5p$ and Cr $3d$ bands), and this change is compensated by the ligand sites as so to keep the total moment equal to $3$ $\mu_{\rm B}$ in both FM semiconductors CrCl$_3$ and CrI$_3$. $J_{\rm eff}$ may of course depend on the magnetization at the ligand sites. Nevertheless, one should clearly understand how this magnetization is treated and what is the relevant physical picture behind Eq.~(\ref{eqn:constraint}). In this respect, we would like to note that, in order to use the magnetic force theorem, the constraining field should be applied at \emph{all} sites of the system, including the ligand ones. Therefore, Eq.~(\ref{eqn:constraint}) implies that the directions of the ligand spins are also rigidly fixed and there is no conceptual difference between the Cr and ligand spins: both groups of spins are treated as localized in the sense that the rotation of the Cr spins has absolutely no effect on the ligand spins and vice versa. Moreover, Fig.~\ref{fig.mgeometry}a implies that the directions of the ligand spins are fixed to be the same as in the FM state. Certainly, this is an approximation and apparently very crude one.

\par The simplest correction to $J_{\rm eff}^{\rm A}$ can be obtained by noting that, besides the interaction between the Cr spins, it also includes interaction with the FM background of the ligand spins. The latter can be corrected by considering the ``ferromagnetic'' rotations, where the spins in both Cr sublattices are rotated by the same angle $\theta$ (Fig.~\ref{fig.mgeometry}b). The corresponding parameter $J_{\rm eff}^{\rm F} = \frac{1}{3} \frac{\partial^2 E(\theta,\theta)}{\partial (2 \theta)^2}$ describes the interaction with the FM background of the ligand spins, and the proper coupling between Cr spins can be found as $J_{\rm eff} = J_{\rm eff}^{\rm A}-J_{\rm eff}^{\rm F}$. All these parameters are listed in Table~\ref{tab:J1c}. One can see that the interaction $J_{\rm eff}^{\rm F}$ is indeed very strong and substantially reduces the total value of $J_{\rm eff}$. Nevertheless, the absolute values of $J_{\rm eff} \sim 100$ meV are still much too high from the viewpoint of magnetic excitation energies, meaning that we can be facing another serious problem and the magnetic excitations described by Eq.~(\ref{eqn:constraint}) may be not the lowest energy ones.

\subsection{\label{sec:constraint} Constrained calculations with the external field}
\par Then, what is wrong and what is missing in the corrected Green's function theory, described by Eq.~(\ref{eqn:constraint})? In this respect, it is important to recall that the theory of spin dynamics, underlying Eqs.~(\ref{eqn:Jij}) and (\ref{eqn:constraint}), is based on the so-called adiabaticity concept~\cite{spindynamics1,spindynamics2}, which states that one can distinguish two types of variables, slow magnetic and fast electronic, so that for each instantaneous magnetic configuration the electronic variables have sufficient time to relax and reach equilibrium. In practical terms, this means that for each constraining field, which controls the direction of magnetization, one should solve self-consistently the set of Kohn-Sham equations and calculate the total energies, which then can be used for the description of low-energy magnetic excitations. This property is implied in derivation of Eqs.~(\ref{eqn:Jij}) and (\ref{eqn:constraint}), where the total energies are additionally replaced by the sum of Kohn-Sham energies, owing to magnetic force theorem. Then, it is clear that the magnetization at the Cr sites, to a good approximation, can be treated as ``slow'' variable. However, what is the nature of the ligand states? Should they be treated as ``slow'' magnetic or ``fast'' electronic variables? The question is rather nontrivial. On the one hand, the ligand magnetization is induced by the Cr one by means of the $dp$ hybridization. From this point of view, it can be viewed as an ``electronic'' variable. On the other hand, since it is a part of the total magnetization, it is always tempting to treat it on the same footing as its Cr counterpart, i.e. as the ``magnetic'' variable. This is precisely what was done in Eq.~(\ref{eqn:constraint}), where we rigidly constrained the direction of magnetization at \emph{all} sites of the system, including the ligand ones. Furthermore, $\vec{\boldsymbol{h}}$ in Eq.~(\ref{eqn:constraint}) has a \emph{matrix} form in the subspace of orbital variables, which is required in order to constrain similar matrix of the magnetization density $\vec{\boldsymbol{m}}$. This is certainly a very strong constraint imposed on the magnetic system. In reality, however, we do not need to constrain all matrix elements of $\vec{\boldsymbol{m}}$. What we need is to fix only the directions of magnetic moment, $\boldsymbol{M}_{i} = {\rm Tr}_L \left\{ \hat{\boldsymbol{m}}_{i} \right\}$, while other constraints can be released. This is also related to how we view the adiabaticity concept. Eq.~(\ref{eqn:constraint}) implies that \emph{all} variables of $\vec{\boldsymbol{m}}$ are slow and magnetic, while the reality can be different: only $\boldsymbol{e}_{i} = \boldsymbol{M}_{i}/|\boldsymbol{M}_{i}|$ are slow variables, while other elements of $\vec{\boldsymbol{m}}$ are fast and have sufficient time to adjust an instantaneous change of $\boldsymbol{e}_{i}$.

\par Therefore, as the next step, we explore the opposite strategy by assuming that all magnetic variables are restricted by the Cr sites, while the ligand magnetization simply follows the Cr one as any other electronic variable. Namely, like in the constraint formalism, we rotate the Cr spins by applying the external field $\vec{\boldsymbol{h}} = (\vec{h}^{x},0,0)$. The rotated configuration of the Cr spins is the same as in Fig.~\ref{fig.mgeometry}a. However, now this field acts only at the Cr sites. Furthermore, each $\hat{\boldsymbol{h}}_{i}$ is constant in the space of orbital variables, $\boldsymbol{h}_{i}^{ab} = \boldsymbol{h}_{i} \delta^{ab}$ ($\delta^{ad}$ being Kronecker delta), and acts only at $\boldsymbol{M}_{i}$, while other degrees of freedom are found from the self-consistent solution of Kohn-Sham equations. In principle, this is well known numerical technique, which is widely used in electronic structure calculations~\cite{Stocks}. In this work, we propose an analytical solution of this problem based on the self-consistent linear response theory~\cite{SCLR}, which allows us to find the self-consistent xc field $\vec{\boldsymbol{b}}$ to first order in $\vec{\boldsymbol{h}}$, the corresponding magnetization $\vec{\boldsymbol{m}}$, and the total energy to second order in $\vec{\boldsymbol{h}}$, which are sufficient to evaluate $J_{\rm eff}$.

\par First, we approximate the xc energy in LSDA (GGA) as~\cite{Gunnarsson}:
\noindent
\begin{equation}
E_{\rm xc} = -\frac{1}{4} \sum_{i} {\rm Tr}_L \left\{  \hat{\boldsymbol{m}}_{i} \hat{I}_{i} \hat{\boldsymbol{m}}_{i} \right\},
\label{eqn:Exc}
\end{equation}
\noindent where $\hat{I}_{i}$ is taken in the matrix form in order to describe asphericity of the xc potential. The corresponding xc field, $\hat{\boldsymbol{b}}_{i} = 2 \delta E_{\rm xc}/\delta \hat{\boldsymbol{m}}_{i}$, at site $i$ is given by
\noindent
\begin{equation}
\hat{\boldsymbol{b}}_{i} = -\frac{1}{2} \left( \hat{\boldsymbol{m}}_{i} \hat{I}_{i} + \hat{I}_{i} \hat{\boldsymbol{m}}_{i} \right) .
\label{eqn:bxc}
\end{equation}
\noindent By applying this equation for $\hat{b}_{i}^{z}$ and $\hat{m}_{i}^{z}$, obtained in collinear calculations for the FM state, one can find the Stoner matrix $\hat{I}_{i}$ (see Appendix~\ref{sec:AppendixB}). We need this step in order to include self-consistency effects for noncollinear magnetic structures, like the one shown in Fig.~\ref{fig.mgeometry}a. Namely, by introducing the rank-4 tensor ${\mathbb I} = \left[ {\mathcal I}^{abcd} \right]$ with ${\mathcal I}^{abcd} = -\frac{1}{2} \left( I^{bc} \delta^{ad} + I^{ad} \delta^{bc} \right)$, Eq.~(\ref{eqn:bxc}) can be rewritten in the compact form $\vec{\boldsymbol{b}} = {\mathbb I} \vec{\boldsymbol{m}}$, which implies summation over last two indices of ${\mathbb I}$ and both indices of $\vec{\boldsymbol{m}}$. Then, the change of the xc field caused by $\vec{\boldsymbol{h}}$ can be found as~\cite{SCLR}:
\noindent
\begin{equation}
\delta \vec{b}^{\alpha} = \left( \left[ 1 - {\mathbb I} {\mathbb R}^{\alpha} \right]^{-1} - 1 \right) \vec{h}^{\alpha},
\label{eqn:bxclr}
\end{equation}
\noindent where ${\mathbb R}^{\alpha}$ is related to spin-dependent elements of the response tensor as ${\mathbb R}^{x} = {\mathbb R}^{y} = \frac{1}{2} \left( {\mathcal R}^{\uparrow \downarrow} + {\mathcal R}^{\downarrow \uparrow} \right)$ and ${\mathbb R}^{z} = \frac{1}{2} \left( {\mathcal R}^{\uparrow \uparrow} - {\mathcal R}^{\downarrow \downarrow} \right)$~\cite{SCLR}. Then, using $\vec{\boldsymbol{h}}$ and $\delta \vec{\boldsymbol{b}}$, one can find the corresponding magnetization change as $\delta \vec{m}^{\alpha} = {\mathbb R}^{\alpha} ( \vec{h}^{\alpha} + \delta \vec{b}^{\alpha} )$ and evaluate the constrained energy (i.e., including the penalty term) to 2nd order in $\vec{\boldsymbol{h}}$ as~\cite{SCLR}:
\noindent
\begin{equation}
\delta E = - \frac{1}{4} {\rm Tr}_L \{ \vec{\boldsymbol{h}} \cdot \delta \vec{\boldsymbol{m}} \}.
\label{eqn:Elr}
\end{equation}
Finally, using $\delta \vec{\boldsymbol{m}}$ one can find the canting angle $\theta$ and evaluate the effective coupling constant $J_{\rm eff}$ from Eq.~(\ref{eqn:jeff}). The obtained energies are also shown in Fig.~\ref{fig.e-theta}, which reveals an excellent quadratic dependence $\delta E = 6J_{\rm eff} \theta^2$. The corresponding parameters $J_{\rm eff}$ are summarized in Table~\ref{tab:J1f}.
\noindent
\begin{table}[h!]
\caption{Parameters of effective exchange coupling (in meV) as obtained using QE (MLWF) method and Eq.~(\ref{eqn:Elr}) for the constrained energies obtained as self-consistent response to the external field applied at the Cr sites. }
\label{tab:J1f}
\begin{ruledtabular}
\begin{tabular}{lc}
          & $J_{\rm eff}$  \\
\hline
CrCl$_3$  & $27.71$   \\
CrI$_3$   & $41.55$   \\
\end{tabular}
\end{ruledtabular}
\end{table}
\noindent One can clearly see that the new values of $J_{\rm eff}$ are substantially reduced and become comparable with the ones derived from the total energy difference between the FM and AFM states. Furthermore, $J_{\rm eff}$ is larger in CrI$_3$, being again in total agreement with the trend obtained in the total energy calculations. We would also like to emphasize that this $J_{\rm eff}$ characterizes the local stability of the FM state. It should be comparable with the total energy difference between AFM and FM states but does not necessarily need to reproduce this difference exactly. The latter may include some other effects, also related to the magnetic polarization of the ligand states, which can be very different in the AFM and FM structures.

\section{\label{sec:conc}Conclusions}
\par Microscopic origin of FM coupling in quasi-2D van der Waals compounds Cr$X_3$ (where $X=$ Cl and I) was investigated on the basis of \emph{ab initio} electronic structure calculations within density functional theory. Although FM coupling in CrCl$_3$ and CrI$_3$ is formally expected from GKA rules for the nearly $90^\circ$ Cr-$X$-Cr exchange path, the realization of this rule in DFT calculations is somewhat nontrivial, which requires special attention in the selection of practical methods and approximation for calculations of interatomic exchange coupling. In the considered Cr trihalides, the ligand states play a crucial role in the origin of FM coupling. However, the value of the exchange coupling strongly depends on ``philosophy'' of how these ligand states should be treated, which is traced back to fundamentals of adiabatic spin dynamics. To certain extend, the exchange coupling depends on the approximation, which is used for the exchange-correlation potential and whether it is treated on the level of LSDA or GGA. We have revealed the microscopic origin of this difference, which is mainly related to the intraatomic Cr $3d$ level splitting. More importantly, this coupling depends on the approximations used for the ligand states. Depending on the approximation, each of which has certain logic behind, the interatomic exchange coupling can vary from weakly antiferromagnetic to strongly ferromagnetic. The best considered approximation, which is consistent with brute force total energy calculations, is to explicitly include the effect of ligand states, but to treat them as electronic degrees of freedom and allow to relax for each instantaneous change of the directions of Cr spins. This strategy can be realized in constrained total energy calculations with the external magnetic field, which can be efficiently treated in the framework of self-consistent linear response theory.

\appendix
\section{\label{sec:AppendixA}
$d$ model decorated with on-site Coulomb interactions}
\par
\begin{figure}[tbp]
\begin{center}
\includegraphics[width=7cm]{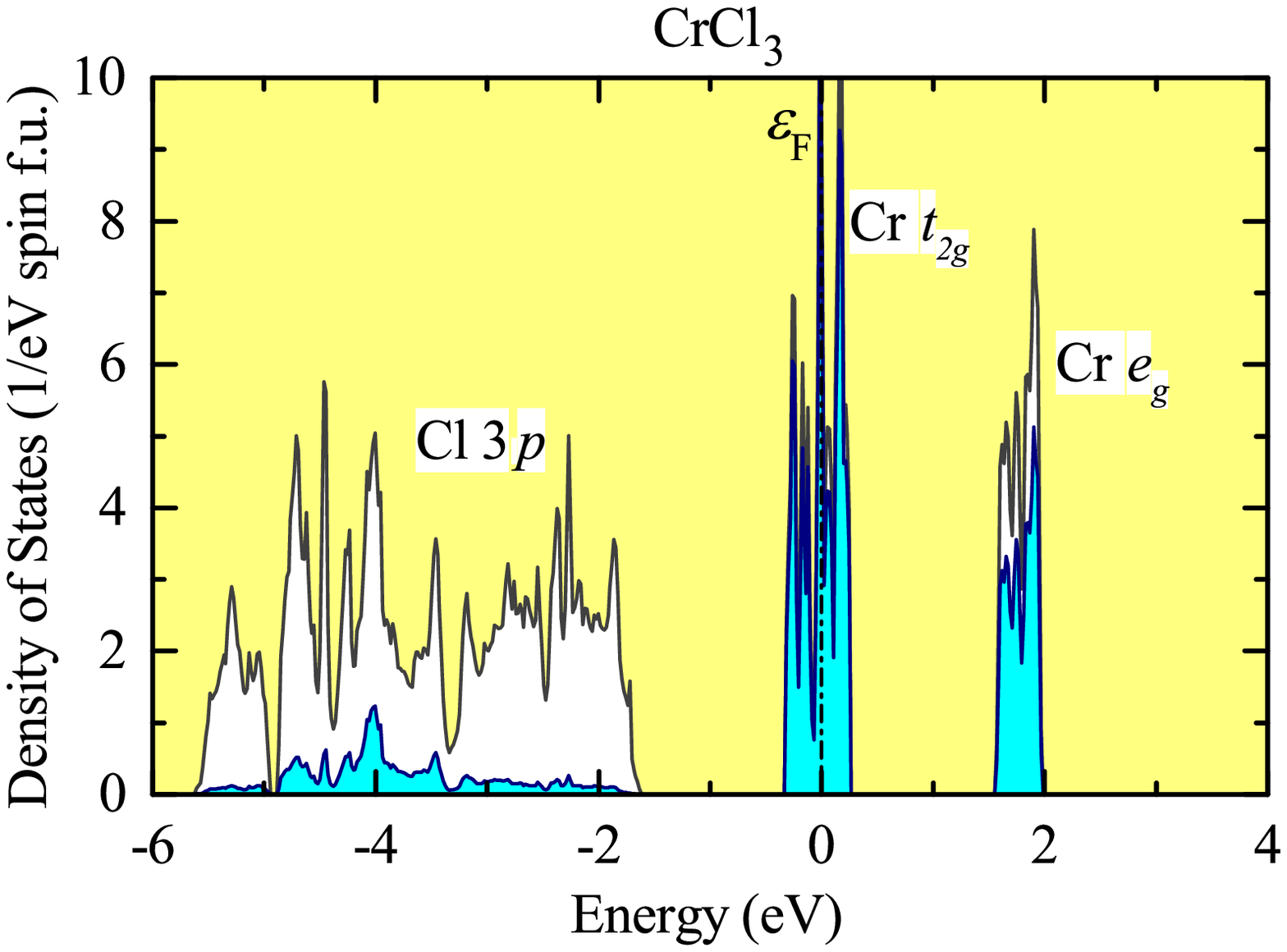}
\includegraphics[width=7cm]{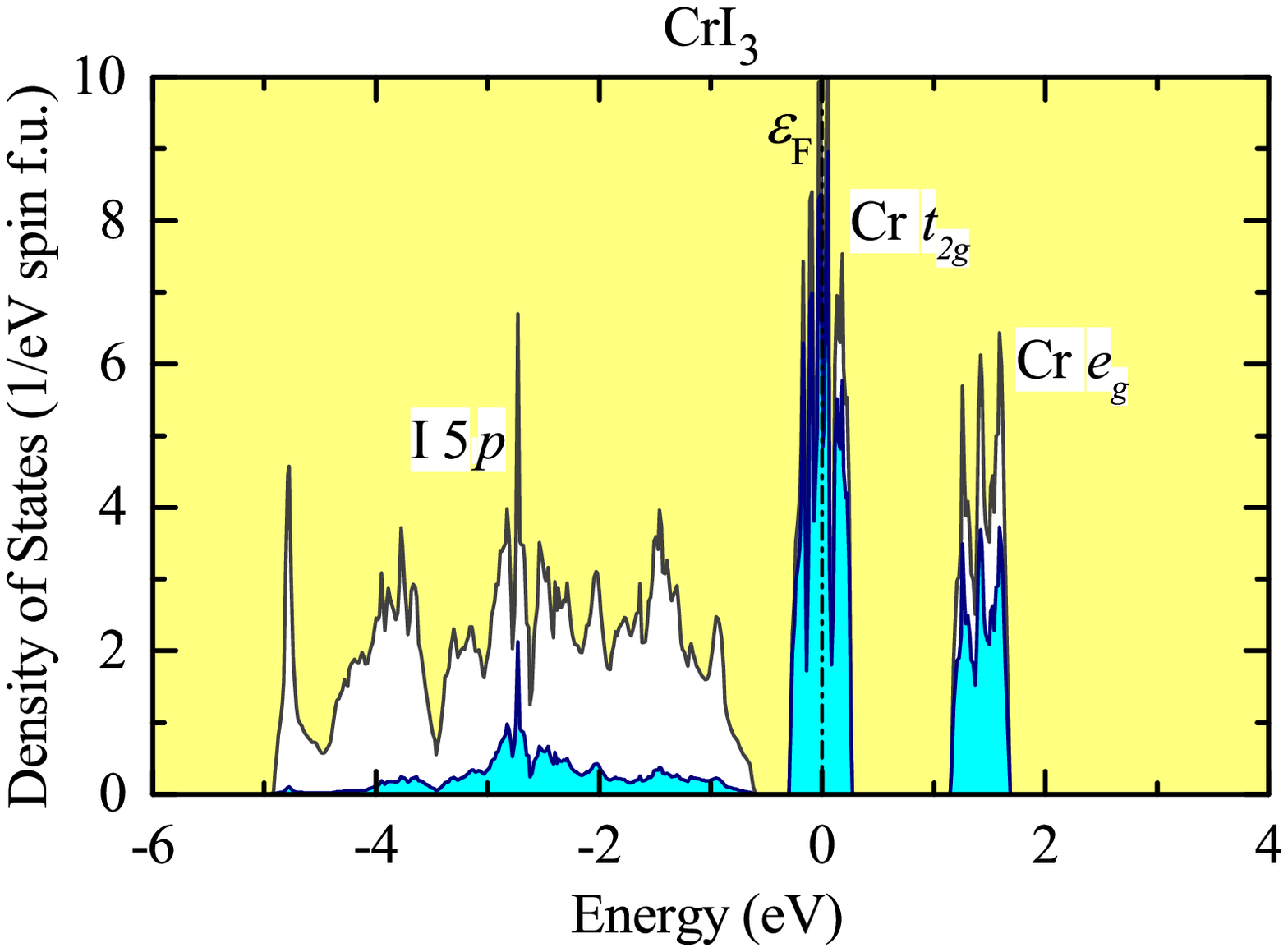}
\end{center}
\caption{(Color online)
Total and partial densities of states of CrCl$_3$ (left) and CrI$_3$ (right) in the local density approximation. The shaded area shows contributions of the Cr $3d$ states. Positions of the main bands are indicated by symbols. The Fermi level is at zero energy (shown by dot-dashed line).}
\label{fig.DOSnm}
\end{figure}

 In this appendix we explore the effect of the on-site Coulomb and exchange interactions on the interatomic exchange coupling, following the general strategy of construction and solution of the effective low-energy model for the Cr $3d$ bands,
\noindent
\begin{equation}
\hat{\cal{H}}  =  \sum_{ij} \sum_{\sigma} \sum_{ab}
H_{ij}^{ab} \hat{c}^\dagger_{i a \sigma} \hat{c}^{\phantom{\dagger}}_{j b \sigma} +
  \frac{1}{2}
\sum_{i}  \sum_{\sigma \sigma'} \sum_{abcd} U_{abcd}
\hat{c}^\dagger_{i a \sigma} \hat{c}^\dagger_{i c \sigma'}
\hat{c}^{\phantom{\dagger}}_{i b \sigma}
\hat{c}^{\phantom{\dagger}}_{i d \sigma'},
\label{eqn.ManyBodyH}
\end{equation}
\noindent constructed in the basis of Wannier orbitals~\cite{review2008}, where $\hat{c}^{\phantom{\dagger}}_{i a \sigma}$ ($\hat{c}^\dagger_{i a \sigma}$) stands for the creation (annihilation) of an electron in the Wannier orbital $a$ at site $i$ with spin $\sigma$. We start with a non-magnetic band structure, obtained in the LMTO method in the local density approximation (LDA). Since Cr $3d$ bands are well isolated from the ligand bands, both in CrCl$_3$ and CrI$_3$ (see Fig.~\ref{fig.DOSnm}), the construction of such model is rather straightforward.
\noindent As in Se.~\ref{sec:Method}, the one-electron parameters $H_{ij}^{ab}$ are identified with the matrix elements of LDA Hamiltonian in the Wannier basis. The only difference is that, since now we are dealing with the non-magnetic band structure, these parameters do not depend on spin indices. The parameters of screened Coulomb interactions, $U_{abcd}$, have been evaluated in the framework of constrained random-phase approximation (cRPA), as described in Ref.~\cite{review2008}. The matrix $\hat{U} = \left[ U_{abcd} \right]$ can be fitted in terms of the on-site Coulomb repulsion $U = F^{0}$, the intraatomic exchange interaction $J = (F^2 + F^4)/14$, and ``nonsphericity'' $B = (9F^2 - 5F^4)/441$ ($F^0$, $F^2$, and $F^4$ being radial Slater's integrals), responsible for the charge stability, and first and second Hund's rules, respectively. This fitting yields (in eV): $U=1.79$ ($1.15$) , $J=0.85$ ($0.78$), and $B=0.09$ ($0.07$) for CrCl$_3$ (CrI$_3$). We note that the screened $U$ is not particularly large. Furthermore, $U$ is smaller in CrI$_3$ due to proximity of the I $5d$ band and, therefore, more efficient screening of Coulomb interactions in the Cr $3d$ band by the ligand $p$ band~\cite{review2008}.

\par Then, the model (\ref{eqn.ManyBodyH}) can be solved in the mean-field Hartree-Fock approximation and the exchange parameters evaluated using Eq.~(\ref{eqn:Jij})~\cite{review2008}. Alternatively, one can use superexchange theory~\cite{PWA,KugelKhomskii}, which yields similar parameters $J_{i}$. The results are summarized in Table~\ref{tab:JiU}.
\noindent
\begin{table}[h!]
\caption{Parameters of exchange interaction (in meV) obtained from solution of model (\ref{eqn.ManyBodyH}) in the mean-field Hartree-Fock approximation. Notations of exchange interactions are shown in Fig.~\ref{fig.Ji}.}
\label{tab:JiU}
\begin{ruledtabular}
\begin{tabular}{lcccccc}
          & $J_{1}$           & $J_{2}$ & $J_{3}$ & $J_{4}$           & $J_{5}$            & $J_{6}$ \\
\hline
CrCl$_3$  & $-0.97$           & $-0.14$ & $-0.26$ & $-0.02$           & $-0.03$            & $-0.16$ \\
CrI$_3$   & $\phantom{-}1.12$ & $-0.27$ & $-0.36$ & $\phantom{-}0.02$ & $\phantom{-}0.07$  & $-0.45$ \\
\end{tabular}
\end{ruledtabular}
\end{table}
\noindent In CrCl$_3$ all interactions are antiferromagnetic. Therefore, the FM order is clearly unstable, both in and between the $xy$ planes. At the first sight, the situation in CrI$_3$ looks somewhat better: at least, the nn coupling $J_{1}$ remains ferromagnetic. However, closer analysis shows that $J_{1}$ is counterbalanced by other AFM interactions, particularly $J_{3}$ and $J_{6}$ in the $xy$ plane as well as $J_{2}$ between the plane. For instance, the parameter $\sum_{i} J_{i}$, which is the measure of Curie temperature in the mean-field approximation and also the Curie-Weiss temperature, appears to be negative. Therefore, the FM state is unstable also in CrI$_3$, contrary to the experimental data. This is consistent with qualitative analysis based on Eq.~(\ref{eqn:SE}): the Coulomb $U$ increases the $t_{2g}$-$e_{g}$ level splitting $ \Delta^{\uparrow \uparrow}_{te}$ and thus suppresses the FM coupling.

\par Thus, to conclude this Appendix we would like to stress the following: (i) The simple $d$ model (\ref{eqn.ManyBodyH}), incorporating on-site Coulomb and exchange interactions, fails to describe properly the FM ordering in CrI$_3$ (and FM ordering in the $xy$ plane of CrCl$_3$). The proper model should include the effects of polarization of the ligand states; (ii) The screened Coulomb interaction $U$ is relatively small in CrCl$_3$ and particularly in CrI$_3$. Thus, plain LSDA (GGA) should provide a good starting point for the analysis of magnetic properties of these compounds, at least on a semi-quantitative level. To certain extent, large $t_{2g}$-$e_{g}$ splitting in QE calculations, based on GGA, mimics the effect of small on-site Coulomb repulsion $U$. A similar conclusion has been reached recently for CrO$_2$, which is another important FM compound, in the joint experimental-theoretical study combining soft-x-ray angle-resolved photoemission spectroscopy and first-principles electronic structure calculations~\cite{CrO2ARPES}.

\section{\label{sec:AppendixB} Calculation of Stoner Matrix}
\par In this Appendix, we briefly discuss the solution of matrix equation,
\noindent
\begin{equation}
\hat{b} = -\frac{1}{2} \left( \hat{m} \hat{I} + \hat{I} \hat{m} \right),
\label{eqn:bxcz}
\end{equation}
\noindent for the matrix $\hat{I}$. Here, $\hat{b}$ and $\hat{m}$ stand for the $z$-component of, respectively, the exchange field and magnetization density, derived from LSDA (GGA) calculations. For simplicity, we drop the site index $i$.

\par The basic idea is to use the representation that diagonalizes $\hat{m}$: $\hat{\tilde{m}} = \hat{O}^{T} \hat{m} \hat{O}$, where $\hat{\tilde{m}} = [\tilde{m}^{a}\delta^{ab}]$. Then, Eq.~(\ref{eqn:bxcz}) can be rewritten as
\noindent
\begin{equation}
\hat{\tilde{b}} = -\frac{1}{2} \left( \hat{\tilde{m}} \hat{\tilde{I}} + \hat{\tilde{I}} \hat{\tilde{m}} \right),
\label{eqn:bxczdiag}
\end{equation}
\noindent where $\hat{\tilde{b}} = \hat{O}^{T} \hat{b} \hat{O}$ and $\hat{\tilde{I}} = \hat{O}^{T} \hat{I} \hat{O}$, for which the matrix elements of $\hat{\tilde{I}}$ can be found as
\noindent
\begin{equation}
\tilde{I}^{ab} = -2  \tilde{b}^{ab}/ ( \tilde{m}^{a} + \tilde{m}^{b} ).
\label{eqn:tIab}
\end{equation}
\noindent Finally, $\hat{I}$ is be obtained from $\hat{\tilde{I}}$ as $\hat{I} = \hat{O} \hat{\tilde{I}} \hat{O}^{T}$.

\par
\begin{figure}[tbp]
\begin{center}
\includegraphics[width=10cm]{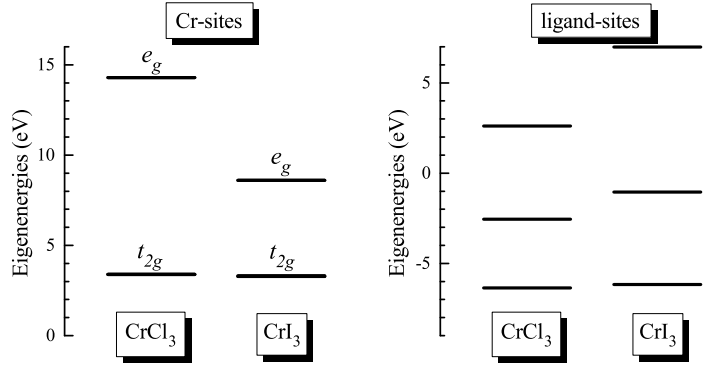}
\end{center}
\caption{
Eigenenergies of Stoner matrices $\hat{I}$.}
\label{fig.Stoner}
\end{figure}

 For further analysis, it is convenient to use the representation, which diagonalizes $\hat{I}$. The results are summarized in Fig.~\ref{fig.Stoner}.
\noindent
\noindent At the Cr-sites, the eigenvalues of $\hat{I}$ split in two groups of levels: threefold nearly degenerate $t_{2g}$ and twofold degenerate $e_{g}$. The $t_{2g}$ levels have practically the same energies for CrCl$_3$ and CrI$_3$ (about $3.3$ eV), while the position of $e_{g}$ levels is different. However, this is to be expected: according to Eq.~(\ref{eqn:tIab}), the matrix elements $\tilde{I}^{aa}$ are inversely proportional to $\tilde{m}^{a}$. The value of $\tilde{m}^{a}$ for the $e_{g}$ states depend on the $dp$ hybridization, which controls the degree of admixture of formally unoccupied Cr $e_{g}$ states into occupied $p$ band of ligands (see Fig.~\ref{fig.DOS}). This admixture is substantially larger in CrI$_3$, which yields larger $\tilde{m}^{a}$ and, therefore, smaller $\tilde{I}^{aa}$ for the $e_{g}$ levels.

\par The averaged value of $\hat{I}$, which enters the regular Stoner model, can be estimated from $b_{\rm av} = I_{\rm av}M$, where $b_{\rm av} = \frac{1}{N}{\rm Tr}_L \{ \hat{b} \}$, $M = {\rm Tr}_L \{ \hat{m} \}$, and $N$ is the number of orbitals ($N=5$ for Cr $3d$ and $N=3$ for ligand $p$ states) as $I_{\rm av} = \frac{1}{N^2} {\rm Tr}_L \{ \hat{I} \}$. For the Cr $3d$ states, it yields $I_{\rm av} = 1.55$ ($1.08$) eV in the case of CrCl$_3$ (CrI$_3$), which is pretty close to typical atomic values of the Stoner parameter for the $3d$ elements~\cite{Gunnarsson}. At the ligand sites, some of the eigenvalues of $\hat{I}$ are negative, so as $I_{\rm av}$. The situation is certainly different from the atomic limit. However, it should be understood that $\hat{b}$ and $\hat{m}$ at the ligand sites have a different microscopic origin: $\hat{b}$ is the difference of \emph{site-diagonal elements} of $\hat{H}^{\uparrow, \downarrow}$ between spins $\uparrow$  and $\downarrow$. This difference is small (see Fig.~\ref{fig.CF}) and does not play a decisive role in the formation of local moments at the ligand sites. These moments are induced by the hybridization with the Cr $3d$ sites and these processes are controlled by \emph{off-diagonal elements} of $\hat{H}^{\uparrow, \downarrow}$ with respect to the site indices. Thus, the direction of $\hat{\boldsymbol{m}}$ at the ligand sites does not necessary coincide with the one of $\hat{\boldsymbol{b}}$, as it follows from the above analysis.


\begin{thebibliography}{99}

\bibitem{CrGeTe3_Nature}
C. Gong, L. Li, Z. Li, H. Ji, A. Stern, Y. Xia, T. Cao, W. Bao, C. Wang, Y. Wang, Z.~Q. Qiu, R.~J. Cava, S.~G. Louie, J. Xia, and X. Zhang,
Nature (London) \textbf{546}, 265 (2017).

\bibitem{CrI3_Nature}
B. Huang, G. Clark, E. Navarro-Moratalla, D.~R. Klein, R. Cheng, K.~L. Seyler, D. Zhong, E. Schmidgall, M.~A. McGuire, D.~H. Cobden, W. Yao, D. Xiao, P. Jarillo-Herrero, and X. Xu,
Nature (London) \textbf{546}, 270 (2017).

\bibitem{MerminWagner}
N.~D. Mermin and H. Wagner,
Phys. Rev. Lett. \textbf{17}, 1133 (1966); \textbf{17}, 1307(E) (1966).

\bibitem{BrunoPRL2001}
P. Bruno, Phys. Rev. Lett. \textbf{87}, 137203 (2001).

\bibitem{CrI3_Lado}
J.~L. Lado and J. Fern\'andez-Rossier,
2D Mater. \textbf{4}, 035002 (2017).

\bibitem{CrI3_Lui}
J. Liu, M. Shi, J. Lu, and M.~P. Anantram,
Phys. Rev. B \textbf{97}, 054416 (2018).

\bibitem{Tsubikawa1960}
I. Tsubokawa, J. Phys. Soc. Jpn. \textbf{15}, 1664 (1960).

\bibitem{Bene1969}
R.~W. Ben\'e, Phys. Rev. \textbf{178}, 497 (1969).

\bibitem{Kanamori_GKA}
J. Kanamori, J. Phys. Chem. Solids \textbf{10}, 97 (1959).

\bibitem{Chaloupka}
J. Chaloupka, G. Jackeli, and G. Khaliullin, Phys. Rev. Lett. \textbf{110}, 097204 (2013).

\bibitem{JHeisenberg}
A.~I. Liechtenstein, M.~I. Katsnelson, V.~P. Antropov, and
V.~A. Gubanov, J. Magn. Magn. Matter. \textbf{67}, 65 (1987).

\bibitem{Stepanov}
E.~A. Stepanov, S. Brener, F. Krien, M. Harland, A.~I. Lichtenstein, and M.~I. Katsnelson,
Phys. Rev. Lett. \textbf{121}, 037204 (2018).

\bibitem{review2008}
I.~V. Solovyev,
J. Phys.: Condens. Matter \textbf{20}, 293201 (2008).

\bibitem{CrO2PRB2015}
I.~V. Solovyev, I.~V. Kashin, V.~V. Mazurenko, Phys. Rev. B \textbf{92}, 144407 (2015).

\bibitem{Heisenberg}
W. Heisenberg,
Zeits. f. Physik \textbf{49}, 619 (1928).

\bibitem{Yosida}
K. Yosida, \textit{Theory of Magnetism}
(Springer-Verlag, Berlin, 1998).

\bibitem{Ku}
W. Ku, H. Rosner, W.~E. Pickett, and R.~T. Scalettar,
Phys. Rev. Lett. \textbf{89}, 167204 (2002).

\bibitem{Mazurenko2007}
V.~V. Mazurenko, S.~L. Skornyakov, A.~V. Kozhevnikov, F. Mila, and V.~I. Anisimov,
Phys. Rev. B \textbf{75}, 224408 (2007).

\bibitem{WannierRevModPhys}
N. Marzari, A.~A. Mostofi, J.~R. Yates, I. Souza, and D. Vanderbilt, Rev. Mod. Phys. {\bf 84}, 1419 (2012).

\bibitem{Oguchi}
T. Oguchi, K. Terakura, and A.~R. Williams,
Phys. Rev. B \textbf{28}, 6443 (1983).

\bibitem{SCLR}
I.~V. Solovyev,
Phys. Rev. B \textbf{90}, 024417 (2014).

\bibitem{CrCl3str}
B. Morosin and A. Narath,
J. Chem. Phys. \textbf{40}, 1958 (1964).

\bibitem{CrI3str}
M.~A. McGuire, H. Dixit, V.~R. Cooper, and B.~C. Sales,
Chemistry of Materials \textbf{27}, 612 (2015).

\bibitem{LMTO}
O.~K. Andersen, Phys. Rev. B \textbf{12}, 3060 (1975).

\bibitem{qe}
P. Giannozzi, S. Baroni, N. Bonini \emph{et. al}, J.Phys.: Condens. Matter \textbf{21}, 395502 (2009).

\bibitem{PBE}
J.~P. Perdew, K. Burke, and M. Ernzerhof, Phys. Rev. Lett. \textbf{77}, 3865 (1996);
\textit{ibid.} \textbf{78}, 1396 (1997).

\bibitem{Wang2011}
H. Wang, V. Eyert, and U. Schwingenschl\"ogl,
J. Phys.: Condens. Matter \textbf{23}, 116003 (2011).

\bibitem{wannier90}
A.~A. Mostofi, J.~R. Yates, G. Pizzi, Y.~S. Lee, I. Souza, D. Vanderbilt, and N. Marzari,
Comput. Phys. Commun. \textbf{185}, 2309 (2014).

\bibitem{Gunnarsson}
O. Gunnarsson,
J. Phys. F: Metal Phys. \textbf{6}, 587 (1976).

\bibitem{lda}
W. Kohn and L.~J. Sham, Phys. Rev. A \textbf{140}, 1133 (1965).

\bibitem{Kanamori}
J. Kanamori, Prog. Theor. Phys. \textbf{17}, 177 (1957).

\bibitem{PRB06}
I.~V. Solovyev, Phys. Rev. B \textbf{73}, 155117 (2006).

\bibitem{PWA} P.~W. Anderson, Phys. Rev. \textbf{115}, 2 (1959).

\bibitem{KugelKhomskii}
K.~I. Kugel and D.~I. Khomskii,
Sov. Phys. Usp. \textbf{25}, 231 (1982).

\bibitem{PRL99}
I.~V. Solovyev and K. Terakura, Phys. Rev. Lett. \textbf{82}, 2959 (1999).

\bibitem{ZSA}
J.~Zaanen, G.~A. Sawatzky, J.~W. Allen, Phys. Rev. Lett. \textbf{55}, 418 (1985).

\bibitem{Heine1}
V. Heine and J.~H. Samson,
J. Phys. F: Metal Phys. \textbf{10}, 2609 (1980).

\bibitem{Heine2}
V. Heine and J.~H. Samson,
J. Phys. F: Metal Phys. \textbf{13}, 2155 (1983).

\bibitem{tyab}
S.~V. Tyablikov, \emph{Methods of Quantum Theory of Magnetism}, Nauka, Moscow, (1975).

\bibitem{PRB98}
I.~V. Solovyev and K. Terakura,
Phys. Rev. B \textbf{58}, 15496 (1998).

\bibitem{Stocks}
G.~M. Stocks, B. Ujfalussy, X. Wang, D.~M.~C. Nicholson, W.~A. Shelton, Y. Wang, A. Canning, and B.~L. Gyorffy, Phil. Mag. \textbf{78}, 665 (1998).

\bibitem{BrunoPRL2003}
P. Bruno, Phys. Rev. Lett. \textbf{90}, 087205 (2003).

\bibitem{spindynamics1}
V.~P. Antropov, M.~I. Katsnelson, B.~N. Harmon, M. van Schilfgaarde, and D. Kusnezov,
Phys. Rev. B \textbf{54}, 1019 (1996).

\bibitem{spindynamics2}
S.~V. Halilov, H. Eschrig, A. Y. Perlov, and P.~M. Oppeneer,
Phys. Rev. B \textbf{58}, 293 (1998).

\bibitem{CrO2ARPES}
F. Bisti, V.~A. Rogalev, M. Karolak, S. Paul, A. Gupta, T. Schmitt, G. G\"untherodt, V. Eyert, G. Sangiovanni, G. Profeta, and V.~N. Strocov,
Phys. Rev. X \textbf{7}, 041067 (2017).

\end{thebibliography}
\end{document}